\documentclass[seceq]{ptptex}
\usepackage{epsfig}
\usepackage{psfrag}
\usepackage{wrapft}

\newcommand{\sub}[1]{\mbox{\tiny{#1}}}
\def\ek{\varepsilon_{\bf k}}

\title{%        %You can use \\ for explicit line-break
Interorbital charge transfers and Fermi-surface deformations in
strongly correlated metals: models, BaVS$_3$ and Na$_{x}$CoO$_2$
}

\author{%       %Use \scshape  for the family name
Frank \textsc{Lechermann}$^{1,2}$, Silke \textsc{Biermann}$^{1}$, and
Antoine \textsc{Georges}$^{1}$
}

\inst{%         %Affiliation, neglected when [addenda] or [errata]
$^1$CPHT {\'E}cole Polytechnique, 91128 Palaiseau Cedex, France\\
$^2$LPT-ENS, 24 Rue Lhomond, 75231 Paris Cedex 05, France}
\abst{%         %this abstract is neglected when [addenda] or [errata]
Fermi-surface deformations in strongly correlated metals, in comparison to
results from band-structure calculations, are investigated.
We show that correlation-induced
interband charge transfers in multi-orbital systems may give rise to
substantial modifications of the actual Fermi surface.
Depending in particular on the relative strength of the crystal-field
splitting and of the Hund's exchange coupling, correlations
may either reinforce orbital polarization or tend to
compensate differences in orbital occupancies, as demonstrated
by investigating a 2-band Hubbard model in the framework of dynamical mean
field theory (DMFT). The physical implications of such interorbital charge
transfers are then explored in two case studies: BaVS$_3$ and 
Na$_x$CoO$_2$.
By means of the DMFT in combination with the local
density approximation (LDA) to density functional theory
(DFT), new insights in the underlying mechanism of the metal-to-insulator
transition (MIT) of BaVS$_3$ are obtained. A strong charge
redistribution in comparison to LDA calculations, i.e., a
depletion of the broader $A_{1g}$ band in favor of the narrower $E_g$
bands just above the MIT is found. In addition, the intriguing problem
of determining the Fermi surface in the strongly correlated
cobaltate system Na$_{x}$CoO$_2$ is discussed.}

\begin{document}

\maketitle

\section{Introduction}
In recent times, theoretical investigations of systems with
strong electronic correlations have become feasible
in realistic settings by using the material-specific
band structure input as a single-particle reference 
system~\cite{Ani97Lic98}. This also raises new questions,
in particular concerning effects stemming from the multi-orbital
character of realistic materials. As most
physical properties of a specific material in the metallic state are
determined by the electronic states close to the Fermi level, one of
the key questions is how these states are affected by strong
Coulomb correlations.
It is a well-documented fact, e.g., in cuprates, that the competition
of localization and itinerancy can result in the appearance of
distinct areas on the Fermi surface with strikingly different physical
properties of the low-energy excitations.
In this work, we concentrate on a more basic aspect, namely
how correlations may lead to a modification of the shape of the
Fermi surface itself, due to interorbital charge transfers.
These modifications may result in a major change of the low-energy 
physics, e.g., by inducing new instabilities.

The track record of density functional theory (DFT) in local (spin)
density or generalized gradient approximation, L(S)DA or
GGA\footnote{If not explicitly noted, it is understood in the
following that in the present context the term 'LDA' covers all
these different approximations to the exchange-correlation energy in
DFT.}, promoted the effective single-particle approach to a standard
tool for electronic structure investigations.
Despite the many successes, this approach is generally
inadequate for strongly correlated systems, and qualitative as well as
quantitative discrepancies are well documented.
However, surprisingly, for many correlated metals,
such as Sr$_2$RuO$_4$~\cite{Lie00}, optimally doped cuprates,
and several heavy-fermion systems~\cite{Run94}, the Fermi surface 
predicted by LDA is apparently in rather good agreement with experimental 
data. Of course, there are also famous counter examples, such as for 
instance the LDA-predicted hole pockets in fcc-Ni which are absent in 
experiment~\cite{Wan73,Yan01}.
Nevertheless, the overall encouraging results for the Fermi surface
obtained by LDA calculations provided 'empirical' confidence that the
Fermi surface is little affected by strong correlations in many materials.
In this paper, we want to draw attention to the fact that there are
indeed cases where electronic correlations do have a significant
influence on the relative orbital occupancies, and on the Fermi-surface 
shape. This may result in a dramatic change of the low-energy physics, 
e.g., by allowing for new nesting possibilities.

Consider first the simple case of a perfectly cubic material
involving a degenerate $t_{2g}$ multiplet. The effective single-particle
(Kohn-Sham) hamiltonian can be diagonalized to yield the bands
$\ek^m$ with orbital index $m$. Assuming a local 
(i.e. ${\bf k}$-independent) self-energy matrix, cubic symmetry implies: 
$\Sigma_{mm'}(\omega)$=$\delta_{mm'}\Sigma(\omega)$.
Hence, the Fermi surface of the interacting system corresponds to the
$k$-points such that: $\ek^m$=$\mu-\Sigma(0)$ (with $\mu$ the chemical
potential), implying that all Fermi-surface sheets are shifted in the 
same manner by correlations.
However, Luttinger's theorem~\cite{Lut60} implies that the total volume
of the Fermi surface is unchanged by interactions, and coincides with the
{\it total} number of electrons. As a result, in this simplest case,
each Fermi surface sheet cannot be changed by correlations and
the relation $\mu-\Sigma(0)$=$\mu_0(n)$ must be satisfied (with $\mu_0$
the chemical potential of the reference system corresponding to the
given total electron number). This teaches us that, for perfect cubic
symmetry and a degenerate multiplet, the Fermi surface can only be changed
by correlations if the self-energy has strong ${\bf k}$-dependence.
In contrast, when orbital degeneracies are lifted due to strong
anisotropies or crystal-field effects, even a local self-energy matrix
can in principle induce Fermi-surface deformations due to its
multi-orbital (non-scalar) structure. The interacting Fermi surface is
now determined by: 
${\rm det}[\mu\delta_{mm'}-H^{mm'}_{\bf k}-\Sigma_{mm'}(0)]$=0.
Even if the off-diagonal components of the self-energy are negligible,
each Fermi-surface sheet can be affected in a different manner according 
to: $\ek^m$=$\mu-\Sigma_{mm}(0)$. Note that Luttinger's theorem does not 
apply 
to each sheet separately, i.e., the {\it partial} orbital occupancies 
$n_m$ do not necessarily correspond to the respective volume of the 
different sheets (even for a diagonal $\Sigma$). However, physical 
intuition suggests that large correlation-induced 
interorbital charge transfers (i.e., changes of the partial $n_m$ for a 
fixed $n$=$\sum_m n_m$), will be associated with strong
Fermi-surface deformations, as confirmed by the studies below.

In this paper, we first explore the relevant processes and mechanisms
that determine interband charge transfers in multi-orbital systems.
Such charge transfers have been recently discussed by several
authors~\cite{Lie00,Pav04,Oka04,Lec05,Ish04,Zho05}. With the vanadium 
sulfide compound BaVS$_3$ and the puzzling sodium cobaltates, we then 
discuss two examples of systems where such repopulations may drive
Fermi-surface deformations that play an essential role for the relevant
physical properties.

\section{Multi-orbital effects in correlated systems \label{modcalc}}
In real materials, the particularities of the band structure
can have substantial influence on the effect of correlations.
Crystal-field splittings and anisotropies giving rise to partial
densities of states with different bandwidths and shapes
lead in general to different renormalizations of the different
bands by the Coulomb interactions.
The net results are charge transfers between different subbands
induced by the Coulomb interactions and a modified Fermi surface
as compared to an effective single-particle reference system,
e.g., as given by the LDA.
The shape of the Fermi surface and the different subband fillings
are in general determined by the interplay of crystal-field splitting,
bandwidth ratios (or more generally the shapes of the partial densities
of states), hybridizations, Coulomb interactions -- in
particular Hund's rule coupling -- as well as the global filling
of the system.
On general grounds, the electrons tend to either avoid the high cost
of Coulomb interactions or to compensate it by enlarging the kinetic
energy. Since the Coulomb interactions are smaller for electrons in
different orbitals, uniform occupations of the subbands are naturally
favored by the multi-orbital interaction vertex.
The difference between the intraorbital Coulomb interaction $U$ and the 
interorbital one $U^{\prime}$ is mathematically related to the Hund's rule
integral $J$ that describes the lowering of the Coulomb interaction by 
exchange between electrons with equal spins.~\footnote{In $t_{2g}$ systems,
e.g., it can be shown that $U-U^{\prime}=2J$ \cite{Cas78Fre97}.}
Thus, a strong Hund's rule coupling will generally lead to rather
uniform partial occupations of the different subbands.
The crystal-field splitting $\Delta$, on the other hand, tends
to orbitally polarize the system. As shown by Manini 
{\it et al.}~\cite{Man02} in the framework of a two-band Hubbard model, 
the relevant energy scale that has to be compared to the crystal-field 
splitting is the {\it renormalized} bandwidth, not the bare one. In a 
system close to the Mott transition, small crystal-field splittings can 
therefore be sufficient to induce a strong orbital polarization.
Recently, this effect was shown~\cite{Pav04} to be crucial to understand 
the insulating oxides LaTiO$_3$ and YTiO$_3$.

The overall orbital charge distribution is thus the result of a
competition between Hund's coupling which tends to level out
differences in the orbital occupations and crystal-field splittings which
favor orbital polarization. This interplay is influenced by the
band structure of the material: in narrow band systems
or for large Coulomb interactions $U$ the crystal-field splittings
are likely to dominate and the system may eventually be driven into
an orbitally polarized state. Moreover, in strongly anisotropic
systems involving bands with very different bandwidths, correlation-induced
band narrowing can be different in the various bands, also
affecting the magnitude of the charge transfer.
The final state is thus a subtle balance between the Hund's rule coupling,
the effective crystal-field splitting and the effective bandwidths of the 
subbands, the latter two renormalized by the Coulomb interactions.

To substantiate these qualitative expectations, we have studied the 
interplay of the crystal-field splitting and Hund's rule coupling as a 
function of $U$ in an anisotropic Hubbard model with two non-hybridizing
bands:
\begin{eqnarray}
\label{hubmod}
\mathcal{H}=&&\sum_{<i,j>m\sigma}\left(t^{(m)}_{ij}-\mu\delta_{ij}\right)
\hat{c}_{im\sigma}^{\dagger}\hat{c}_{jm\sigma}+
\Delta\sum_{i\sigma}(\hat{n}_{i2\sigma}-\hat{n}_{i1\sigma})\\ \nonumber
&&\quad+U\sum_{im}\hat{n}_{im\uparrow}\hat{n}_{im\downarrow}+
U'\sum_{i\sigma}\hat{n}_{i1\sigma}\hat{n}_{i2\bar{\sigma}}+
U''\sum_{i\sigma}\hat{n}_{i1\sigma}\hat{n}_{i2\sigma}\quad,
\end{eqnarray}
where $\hat{c}_{im\sigma}^{\dagger}(\hat{c}_{jm\sigma})$ creates
(annihilates) an electron with spin $\sigma(=\uparrow,\downarrow)$ and
orbital index $m(=1,2)$ at the $i$th site. The interorbital interactions
$U'$,$U''$ are described by combinations of the on-site
Coulomb interaction $U$ and the local Hund's coupling integral $J$. We
chose the parametrization $U'$$=$$U-2J$ and $U''$$=$$U-3J$, proven to
be suitable for t$_{2g}$ substates \cite{Cas78Fre97}. Crystal-field 
effects are taken into account via the energy splitting $2\Delta$ between 
the orbitals. Spin-Flip and pair-hopping terms were neglected in
(\ref{hubmod}).
For the numerical applications we choose the lattice to be the
infinite-connectivity Bethe lattice giving rise to a semi-elliptical
density of states (DOS) of bandwidth $W_m$ for band $m$.

This model is treated within dynamical mean field theory (DMFT):
the lattice model is mapped onto a local-impurity problem in
an effective self-consistent bath. The self-consistency condition
relates the local lattice Green's function to the impurity problem
in the usual manner~\cite{Geo96}. The impurity problem is
solved using the quantum Monte-Carlo (QMC) Hirsch-Fye
algorithm~\cite{Hir86} with up to 128 slices
in imaginary time $\tau$ and $10^6$ sweeps.

\begin{figure}[t]
\parbox[c]{4.5cm}{% Fig mod1a:
\epsfclipon
\epsfig{file=scircle1.eps,width=4.5cm}}
\hspace{-0.1cm}
\parbox[c]{4.5cm}{% Fig mod1b:
\epsfclipon
\epsfig{file=occ-uj.eps,width=4.5cm}}
\hspace{-0.1cm}
\parbox[c]{4.5cm}{% Fig mod1c:
\epsfclipon
\epsfig{file=zj.eps,width=4.5cm}}\\
\hspace*{2.4cm}(a)\hspace{4cm}(b)\hspace{3.8cm}(c)\\[-0.35cm]
\caption{Two-band model calculations with $W_2$=0.5$W_1$,
$n$=1 and $\Delta$=0. (a) Non-interacting DOS. (b) Band fillings for
different $U/J$ series in the metallic regime. (c) Band-resolved QP
residue for different $U/J$ series. Dark (red) lines belong to 
properties of band 1(2).\label{Figmod1}}
\vspace*{0.2cm}
\parbox[c]{4.5cm}{% Fig mod2a:
\epsfclipon
\epsfig{file=scircle3.eps,width=4.5cm}}
\hspace{-0.1cm}
\parbox[c]{4.5cm}{% Fig mod2b:
\epsfclipon
\epsfig{file=occ-un05.eps,width=4.5cm}}
\hspace{-0.1cm}
\parbox[c]{4.5cm}{% Fig mod2c:
\epsfclipon
\epsfig{file=zjincom.eps,width=4.5cm}}\\
\hspace*{2.4cm}(a)\hspace{4cm}(b)\hspace{3.8cm}(c)\\[-0.35cm]
\caption{Two-band model calculations in the incommensurate case with 
$W_2$=0.5$W_1$,
$n$=0.5 and $\Delta$=0. (a) Non-interacting DOS. (b) Band fillings for
$J$=0 and $U/J$=4 series in the metallic regime. (c) Band-resolved QP
residue $Z$ for both series. Dark (red) lines belong to properties of 
band 1(2).\label{Figmod2}}
\end{figure}
\begin{figure}[t]
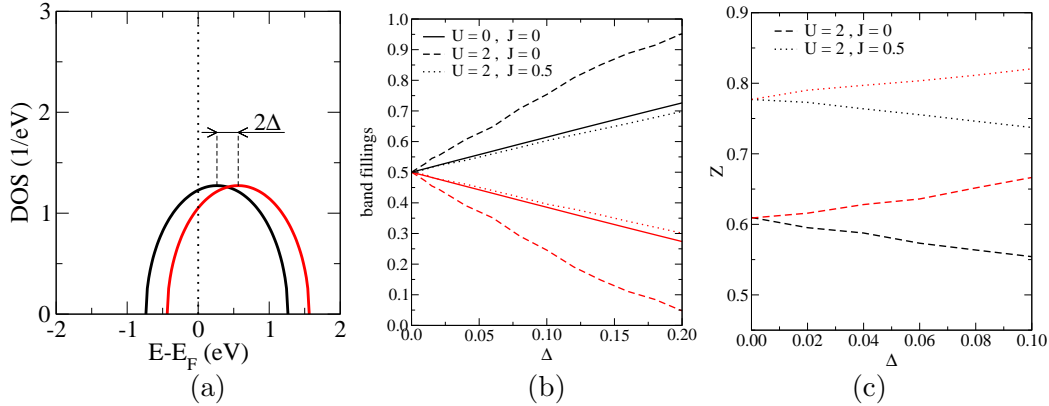

\parbox[c]{4.5cm}{% Fig mod3a:
\epsfclipon
\epsfig{file=scircle2.eps,width=4.5cm}}
\hspace{-0.1cm}
\parbox[c]{4.5cm}{% Fig mod3b:
\epsfclipon
\epsfig{file=occ-ujdelta.eps,width=4.5cm}}
\hspace{-0.1cm}
\parbox[c]{4.5cm}{% Fig mod3c:
\epsfclipon
\epsfig{file=zdelta.eps,width=4.5cm}}\\
\hspace*{2.4cm}(a)\hspace{4cm}(b)\hspace{3.8cm}(c)\\[-0.35cm]
\caption{Two-band model calculations with $W_2$=$W_1$, $n$=1 and
varying crystal field $\Delta$. (a) Non-interacting DOS. (b) Band fillings
for different $U$, $J$ settings in the metallic regime. (c) Band-resolved
QP residue $Z$.\label{Figmod3}}
\end{figure}

In the following, $n$ denotes the total filling, and $n_1$,$n_2$ the 
respective subband filling. Figure 1 shows results for the two-band model 
with different bandwidths (ratio $b$=$W_2/W_1$=1/2) in the absence of 
crystal-field splitting. The overall
filling of one electron ($n$=1) is kept constant during the calculation,
but the partial fillings change from their non-interacting values of
$n_1$$\sim$2/3 and $n_2$$\sim$1/3. Figure 1a visualizes the non-interacting
densities of states (DOS), while Fig. 1b displays the partial occupations
of the two bands as a function of the Coulomb interaction $U$ for
different values of $U/J$. The overall tendency to equalization of
the orbital occupancies with increasing Coulomb interaction and in
particular close to the insulating regime at large $U$ is a
consequence of the bandwidths differences being less important
the stronger the renormalizations become. We note that the
quasiparticle residues $Z_m$, as extracted from the linear regime
of the imaginary part of the self-energies on the Matsubara axis,
are of similar magnitude (even if, interestingly the broad band
is slightly stronger correlated, due to its bigger occupation).
The influence of $J$ which reinforces the orbital compensation is
evident: for decreasing $U/J$ ratios the orbital compensation effect is
substantially more pronounced.

For small $J$ -- the effect is striking at $J$=0 but also
present for selected $U/J$ -- a small-$U$ regime is observed
where increasing mutual electronic interactions first lead to further
orbital polarization. This should stem from the fact that by increasingly
populating the broader band the electrons can gain kinetic energy and
thus may compensate the increased potential energy. To quantify this
subtle balance more rigorously we have considered the above model in a
static approximation which allows for an exact solution. We replace the
operators by their mean values
$\hat{n}_m\rightarrow\langle\hat{n}_m\rangle$$=$$n_m$ and use
a simple Friedel representation for the kinetic energy:
$E_{\sub{kin}}$$\sim$$n(2-n)$.
From a minimization of the sum of the kinetic and on-site potential energy
one obtains the occupations
\begin{equation}
\label{statmod}
n_1=\frac{(1+b(n-1))W_1-nU(1-\frac{5}{a})+2\Delta}
{(1+b)W_1-2U(1-\frac{5}{a})}\quad,\quad n_2=n-n_1\quad.
\end{equation}
with $n$=$n_1+n_2$ and $a$=U/J. Choosing $b$$\le$1 and refering to 
$n_{1,0}$ as the occupation in the non-interacting case, the condition for 
$n_1$$>$$n_{1,0}$ reads:
\begin{eqnarray}
\label{polcond}
1.&&\qquad a>5\quad{\rm for}\quad\Delta>\frac{W_1}{4}(n-2)(1-b)\qquad,
 \nonumber \\
2.&&\qquad a<5\quad{\rm for}\quad\Delta<\frac{W_1}{4}(n-2)(1-b)\qquad.
\nonumber
\end{eqnarray}
Independent of bandwidth, filling and crystal-field splitting, the system will
thus polarize for $U/J$$>$5 and compensate for $U/J$$<$5, with respect to the 
non-interacting case. 
This limiting value can already be obtained while neglecting
the kinetic-energy terms. It follows, e.g., from a Hartree-Fock treatment
of the interaction term alone~\cite{Oka04,Ish04,Zho05}. In fact, it can be
shown that in the static limit the polarization/compensation 
regimes are in most cases independent of the specific form of the band 
energy. In the small-$U$ regime
where fluctuations are weak and the bandwidth renormalizations small,
the static limit indeed describes the charge flow correctly
(see Fig.1). However, for larger $U$ dynamical effects and the
changes in kinetic energy due to the bandwidth renormalizations become
important: in this regime, the system tends to level out orbital occupation
differences and eq. (\ref{statmod}) no longer applies.

The situation is different for non-integer total filling $n$, 
(see Fig.~\ref{Figmod2}). Though for small $U$ the behavior is again
qualitatively correctly described by (\ref{statmod}), the system always 
tends to polarize in the large-$U$ limit. Indeed, in the 
doped case there is no Mott transition and fluctuations are weaker, even 
for large $U$ (Fig.~\ref{Figmod2}c). Interestingly, there appears to be
also a change in the degree of correlation for the different bands when
turning on $J$.

Finally, we present results from calculations for a two-band model with 
finite crystal-field splitting and equal bandwidths $W_1$=$W_2$
in the commensurate case with $n$=1. Increasing Coulomb interactions
drastically increase the orbital polarization, the stronger
the larger the crystal field $\Delta$.
However, a large Hund's coupling $J$ works against this polarization,
as seen for $U/J$=4, which brings the orbital occupations back to
their non-interacting values. As seen in Fig.~\ref{Figmod3}c, the large
differences in the occupations also give rise to substantially
different quasiparticle residues $Z_m$.

From these model calculations we conclude that in general
Coulomb interactions in non-degenerate multiband systems
can induce substantial charge transfers between single subbands.
Hund's rule coupling generally leads to a compensation of
orbital occupation differences, whereas crystal-field splittings
tend to polarize the system. The net charge flow depends on
the fine balance of these effects, renormalized by the Coulomb
interactions and influenced by the anisotropies of the system.
This explains, why indeed in some systems such as the series
of $3d^1$ perovskite compounds studied in Ref.~\citen{Pav04}
correlations were found to orbitally polarize the system
whereas in others (e.g. Sr$_2$RuO$_4$~\cite{Lie00,Oka04})
compensation effects were observed.
Below we discuss two further recent examples, in which the
question of orbital polarization or compensation effects
acquires deeper importance by the fact, that the charge
transfers induce substantial Fermi surface modifications:
the first one is the vanadium sulfide compound BaVS$_3$ where the
charge compensation effect was recently proposed \cite{Lec05} to be
at the origin of a Fermi-surface modification that could be
crucial for explaining the low-temperature properties
of this compound. The second example is the thermoelectric
cobaltate system Na$_x$CoO$_2$, where very recent angle-resolved 
photoemission (ARPES) experimental data~\cite{Yan05,Gec05} seem to hint 
towards a charge polarization effect, whereas there are conflicting results 
concerning this issue from theoretical investigations of this 
problem~\cite{Zha04,Ish04,Zho05}.

\section{The metal-insulator transition of BaVS$_3$}
\subsection{Brief overview of the physical properties of BaVS$_3$}
\begin{figure}[t] % Fig1 : BaVS3 crystal strcuture
\parbox[t]{4cm}{(a)\\[0.2cm]
\epsfxsize=0.4\textwidth
\epsfbox{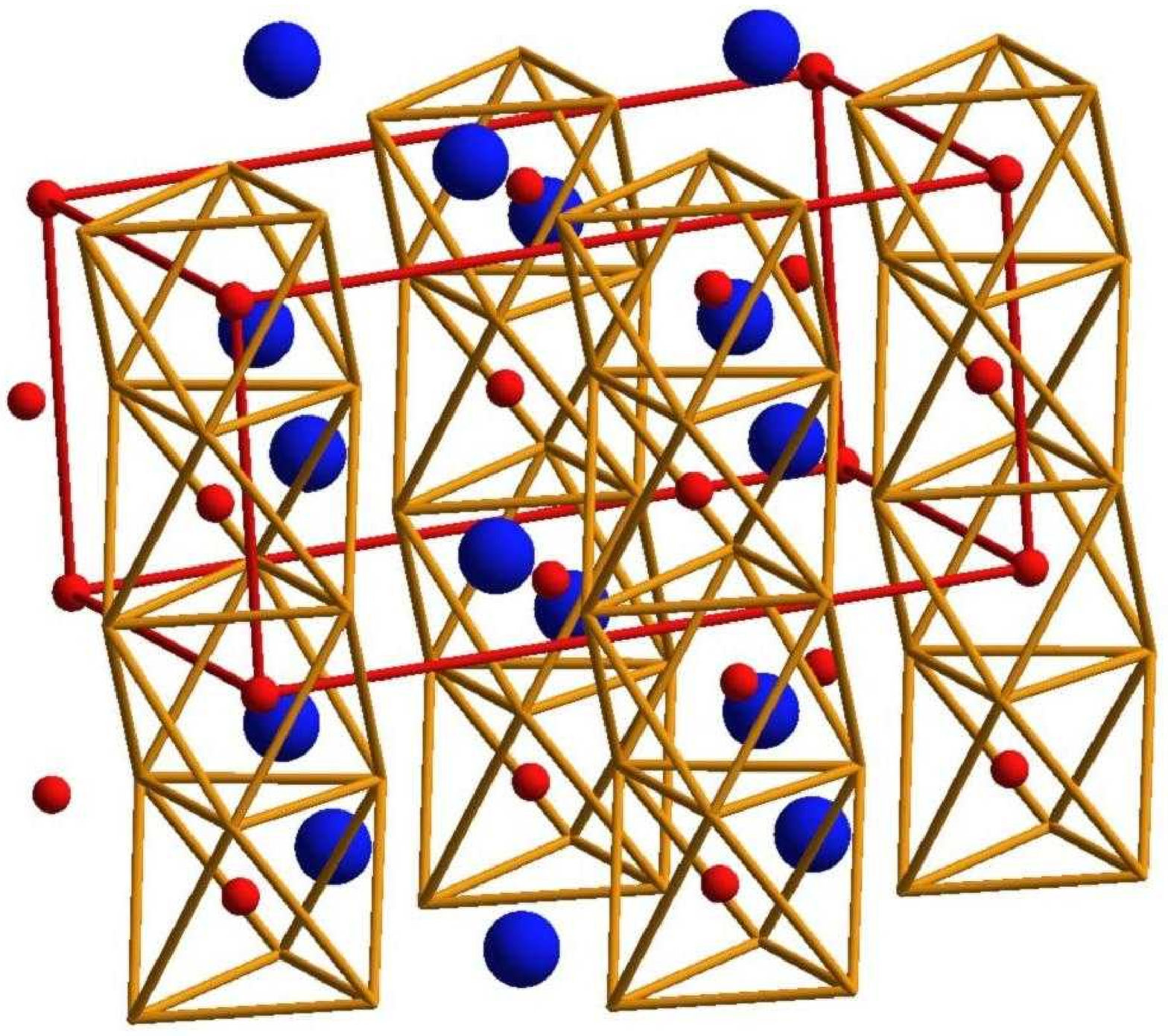}}
\parbox{1.75cm}{\hfill}
\parbox[t]{7cm}{(b)\\[0.2cm]
\epsfxsize=0.4\textwidth
\epsfbox{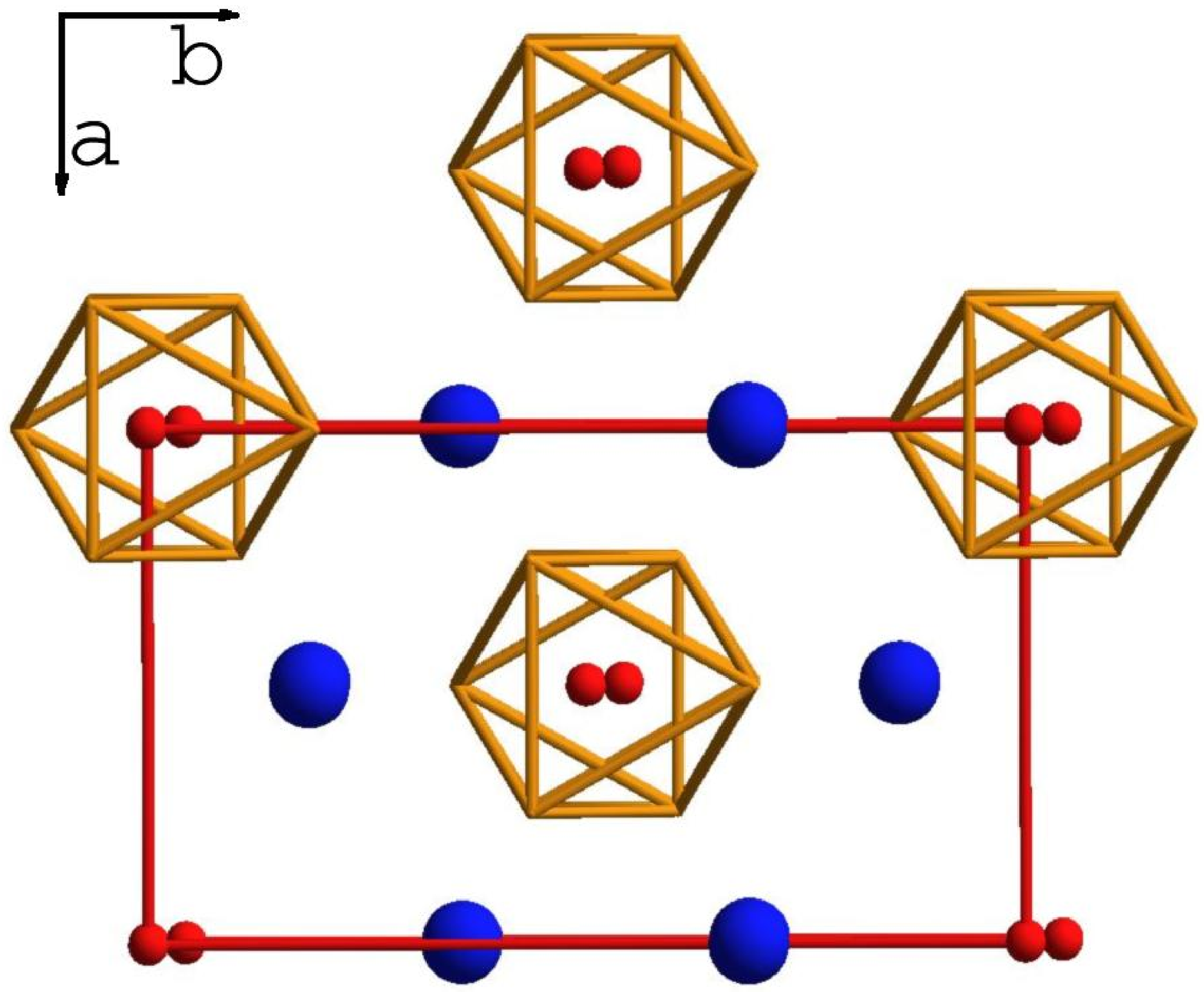}
\vspace{0.2cm}\hspace{-0.1cm}\epsfxsize=0.055\textwidth
\hspace{0.25cm}\epsfbox{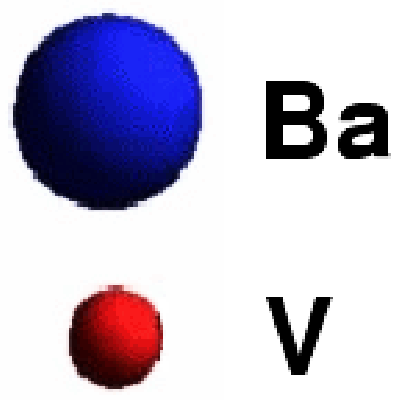}}
\caption{BaVS$_3$ in the orthorhombic ($Cmc2_1$) structure.
(a) 3D view, (b) view along the $c$ axis.}
\label{Fig1}
\end{figure}

\begin{figure}[t]
\parbox{\halftext}{% Fig 2: Resistivity and susceptibility
\epsfclipon
\epsfig{file=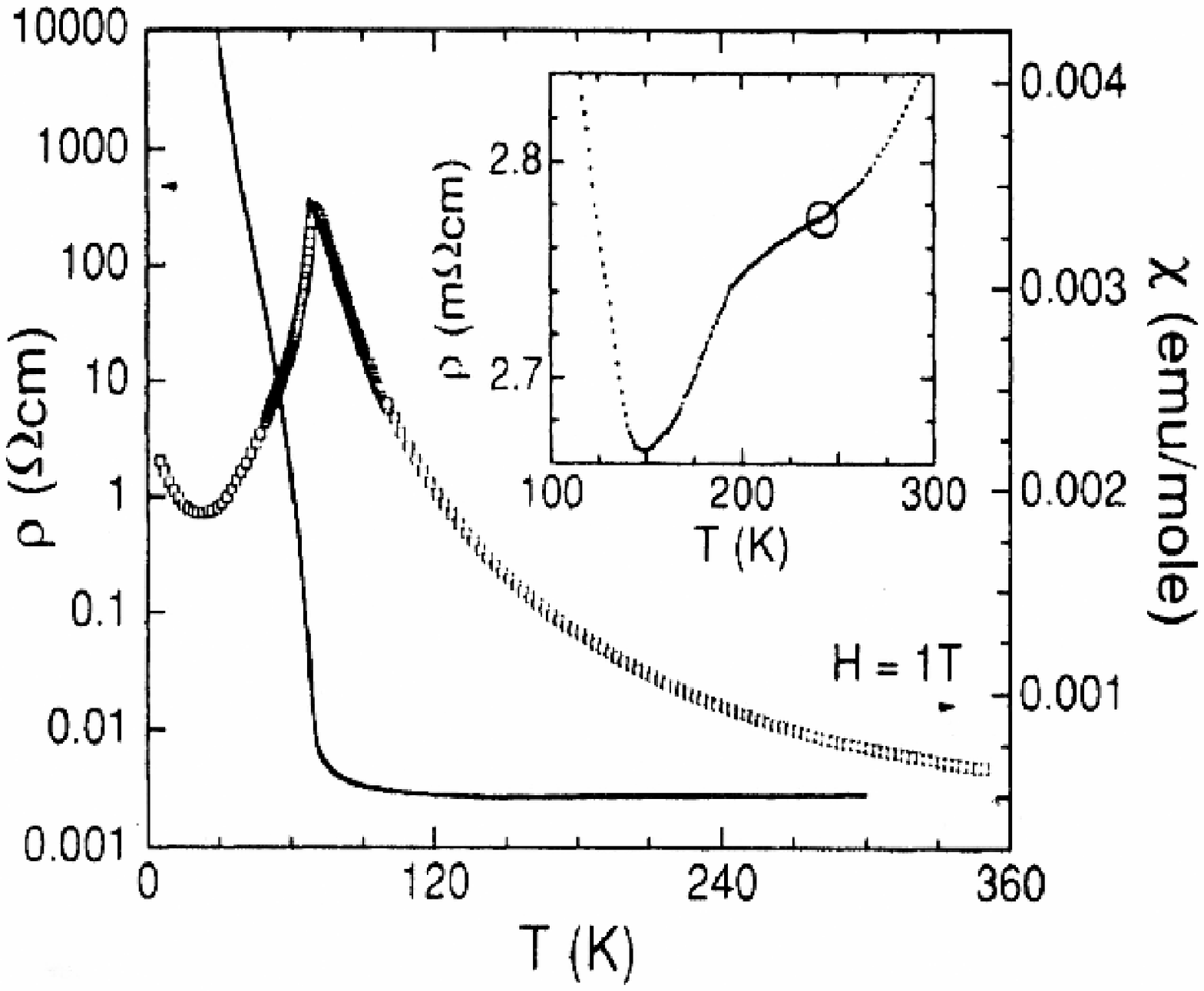,width=6.25cm}
\caption{Electrical resistivity (full line) and magnetic susceptibility
         (circles) of BaVS$_3$ (from Ref.~\citen{Gra95}).}
\label{Fig2}}
\hfill
\parbox{\halftext}{% Fig 3: Hall constant
%\vspace*{-0.5cm}
\epsfclipon
\epsfig{file=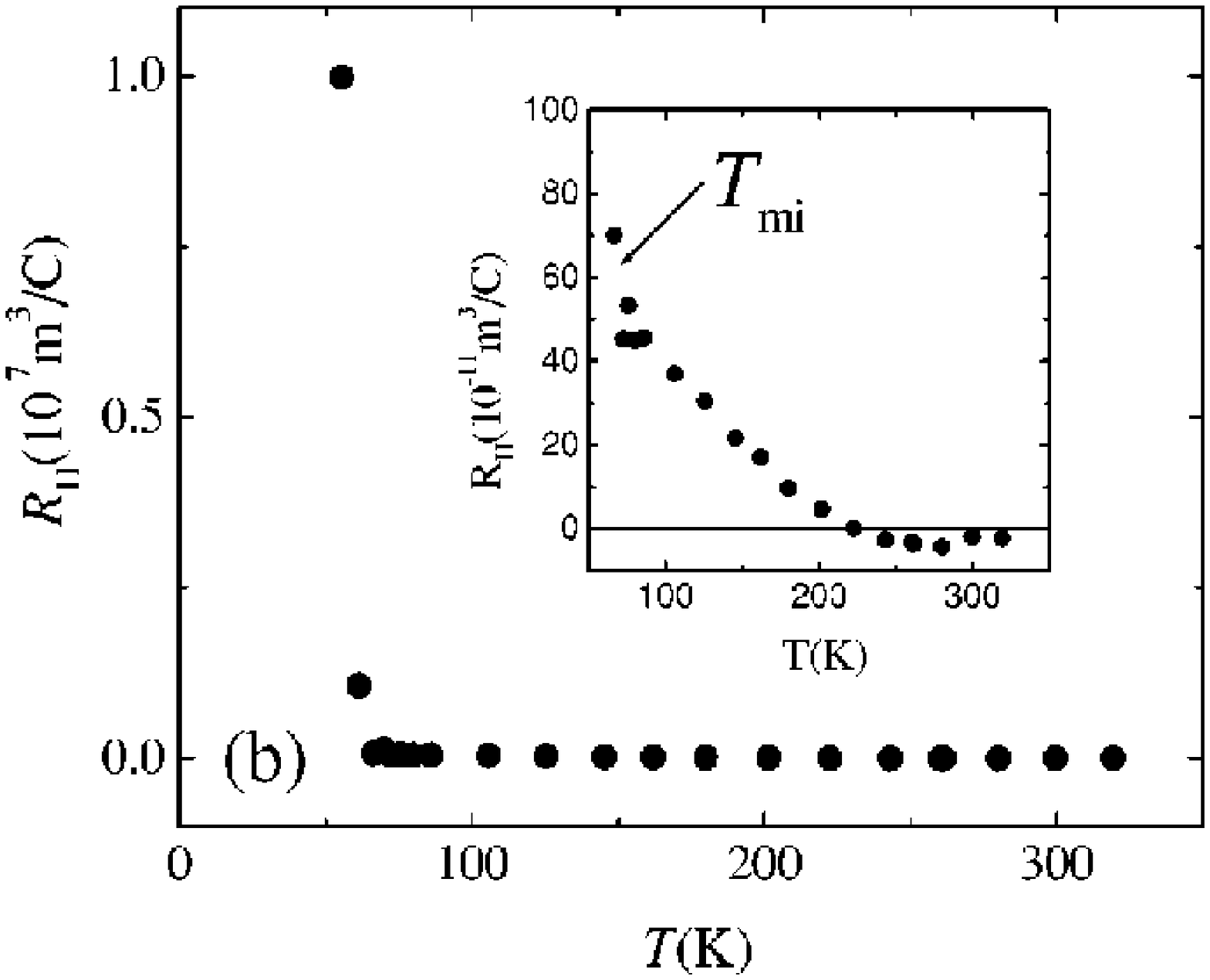,width=6cm}
\caption{Hall coefficient $R_H$ versus temperature. The large increase
close to $T_{\sub{MIT}}$ indicates the carrier loss (from
Ref.~\citen{Boo99}).}
\label{Fig3}}
\end{figure}

BaVS$_3$ is well known as a long-standing challenging compound,
exhibiting unusual electrical and magnetic
properties~\cite{Tak77,Mas79,Mat91,Gra95,Boo99,Wha02}. At room
temperature, this vanadium sulfide crystallizes in a hexagonal
($P6_3/mmc$) structure \cite{Gar69}, whereby straight chains of face
sharing VS$_6$ octahedra are directed along the $c$ axis. Remarkably,
BaVS$_3$ undergoes three distinct continuous phase transition with
decreasing temperature. At $T_{\sub{S}}$$\sim$240 K the crystal structure
transforms into an orthorhombic ($Cmc2_1$) structure (see
Fig.~\ref{Fig1}) \cite{Ghe86}, thereby creating an anisotropy in the
$ab$-plane, i.e., perpendicular to the chain direction, and a zigzag
distortion of the VS$_3$ chains in the $bc$-plane. Additionally, the sign
of the Hall coefficient changes from negative to positive~\cite{Boo99}.
The orthorhombic as well as the hexagonal unit cell both contain two
formula units of BaVS$_3$, whereby the relevant V atoms are equivalent by
symmetry. The system displays a MIT at $T_{\sub{MIT}}$$\sim$70 K into a
paramagnetic phase, hence BaVS$_3$ is an example where a MIT occurs
without intervening magnetic ordering. Apparently, a magnetic transition
finally takes place at $T_{\sub{X}}$$\sim$30 K, where an incommensurate
antiferromagnetic order seems to be established \cite{Nak00}.\\
The paramagnetic conducting phase behaves as metallic ($d\rho/dT$$>$0)
above $\sim$150 K, from where the resistivity~\cite{Gra95} (see
Fig.~\ref{Fig2}) and the Hall coefficient~\cite{Boo99} (see
Fig.~\ref{Fig3}) increase up to the MIT. From diffuse scattering
experiments~\cite{Fag03}, correlated structural fluctuations along the
direction of the VS$_3$ chains were deduced. These fluctuations were
interpreted as a precursor for a commensurate charge density wave (CDW)
instability associated with a possible Peierls mechanism that triggers the
MIT. The obtained wave vector for the structural instability in the
orthorhombic cell $\mathbf{q}_{\sub{MIT}}$=$(1,0,\frac{1}{2})_{O}$
is identical with the one derived by Inami {\sl et al.} \cite{Ina02} which
indeed detected a doubling of the unit cell below $T_{\sub{MIT}}$. In a
recent x-ray study Fagot {\sl et al.}~\cite{Fag04} were able to
characterize the crystal structure below 70~K. The unit cell with a
monoclinic distortion (space group $Cm$) now contains four formula units,
whereby there appears to be a trimerization of the $V$ atoms along the $c$
axis, corresponding to a dominant $2k_F$ distortion~\cite{Hui82}. It should
be kept in mind however that the conduction anisotropy within the system
is not strongly pronounced ($\sigma_c/\sigma_a$$\sim$3-4)~\cite{Mih00},
making pure 1D interpretations questionable. Furthermore, the ``bad
metal'' regime above the MIT with the resistivity minium at $\sim$150 K
is still not quite understood. In ultraviolet
photoelectron-spectroscopic~\cite{Itt91} and
photoemission~\cite{Nak94,Mit05} studies no Fermi edge was observed in
the temperature regime around $T_{\sub{MIT}}$. This raised some
speculation about a possible realization of a Luttinger electron liquid
in this compound, which however seems unrealistic due to the intriguing
interplay of 3D and 1D features. The large increase of the Hall
constant~\cite{Boo99} within the precursive regime seems to suggest
that carriers are scattered into immobile states. In addition, an
important feature of the phase above $T_{\sub{MIT}}$ is the existence of
local moments as revealed by the Curie-Weiss form of the magnetic
susceptibility~\cite{Gra95}. The effective moment corresponds
approximately to one localized spin-$1/2$ per two V sites. Since the
formal valence is V$^{4+}$, corresponding to one electron in the
3$d$-shell, this can be interpreted as the effective localization of
half of the electrons. At $T_{\sub{MIT}}$, the susceptibility rapidly
drops, and the electronic entropy is strongly suppressed~\cite{Ima96}.\\
The electronic structure close to the Fermi level in the orthorhombic
($Cmc2_1$) phase is dominated by low-lying V(3$d$)-$t_{2g}$ states.
Due to symmetry, these states are split into one $A_{1g}$ and two
$E_g$ states per V atom. First-principles DFT-LDA calculations for
$Cmc2_1$-BaVS$_3$\cite{Mat95} do yield a V(3$d$)-S(3$p$) hybridization
which is strong enough to account for the weak anisotropy of the
transport properties. No band-gap opening has been reached within LDA.
Instead, very narrow $E_g$ bands right at the Fermi level, and a nearly filled
dispersive band with mainly $A_{1g}$ character extending along the
$c^*$ direction (${\bf c}^*$ is the reciprocal unit cell vector along the
$c$ axis of the system) have been found. This is consistent with a simple
model proposed early on by Massenet {\sl et al.}~\cite{Mas79}. However,
the occupancy of the
narrow $E_g$ bands found within LDA is too low to account for the
observed local moment in the metallic phase. The nature of the CDW
instability is also left unexplained by the LDA calculations. Indeed, the
calculated norm of the Fermi wave vector of the broad $A_{1g}$ band is
found to be $2k_F^{\sub{LDA}}$$\simeq$$0.94c^*$~\cite{Mat95}, while the
observed wave vector of the instability is
${\mathbf q}_{\sub{MIT}}$=$0.5{\bf c}^*$~\cite{Fag03}. Therefore, the
simple picture of a CDW at $q_{\sub{MIT}}$=$2k_F$ associated only with
the $A_{1g}$ band is untenable within LDA. It is likely that the $E_g$
states also participate in the instability, still the LDA band
structure does not provide a Fermi-surface nesting that is in line
with experimental findings. Hence, ab-initio calculations based on L(S)DA
are not sufficient to explain the complex electronic structure of
BaVS$_3$. Static L(S)DA+U and/or GGA+U schemes appear also not to be
adequate. Despite the fact that a band gap is obtained for
$Cmc2_1$-BaVS$_3$ when enforcing magnetic order, this does not
resemble the experimental scenario of an CDW mechanism underlying the
transition from a Curie-Weiss-like metal into an paramagnetic insulator.\\
In the following we will show that within a DMFT framework, using the
LDA electronic structure as a starting point, one may indeed reconcile 
theory with experimental findings. On the basis of this LDA$+$DMFT 
treatment we propose correlation effects in a multi-orbital context as an
explanation for the discrepancies between band theory predictions and
experiments, in line with the results of our model investigations in 
\S\ref{modcalc}. Specifically, we show that interorbital charge transfers
occur which lower the occupancy of the $A_{1g}$ orbital in favor of the
$E_g$'s. Thus leading to a Fermi surface modification by mainly
shifting $k_F(A_{1g})$ towards lower values. From a calculation of the
local susceptibilities, we demonstrate that local moments are formed in
the metallic phase due to the low quasiparticle coherence scale induced
by the strong correlations (in particular for the narrow $E_g$ bands).
\subsection{Calculational Approach}
\begin{figure}[t] % d-orbitals
\parbox[t]{3.5cm}{\hspace*{2cm}(a)\\[-0.05cm]
\epsfxsize=0.35\textwidth
\epsfbox{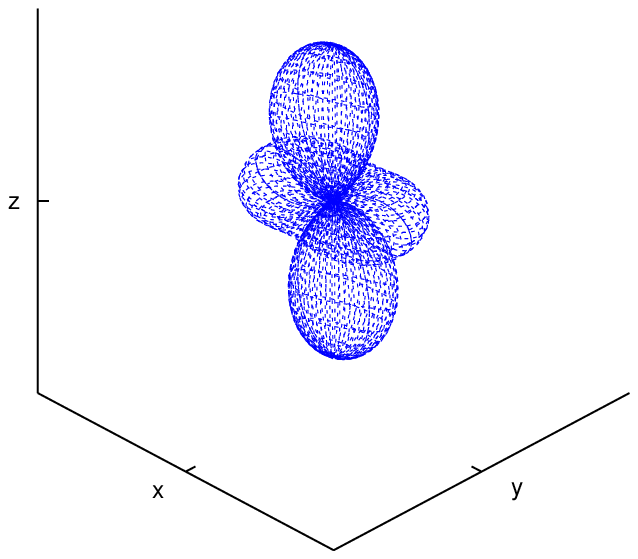}}
\parbox{0.8cm}{\hfill}
\parbox[t]{3.5cm}{\hspace*{2cm}(b)\\[-0.05cm]
\epsfxsize=0.35\textwidth
\epsfbox{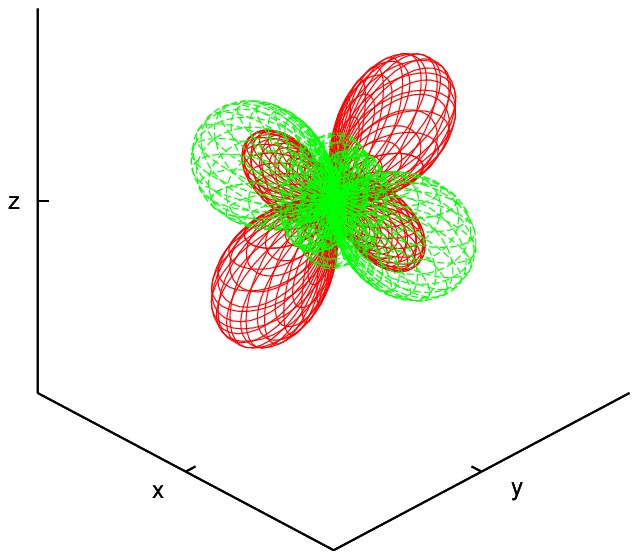}}
\parbox{0.8cm}{\hfill}
\parbox[t]{3.5cm}{\hspace*{2cm}(c)\\[-0.05cm]
\epsfxsize=0.35\textwidth
\epsfbox{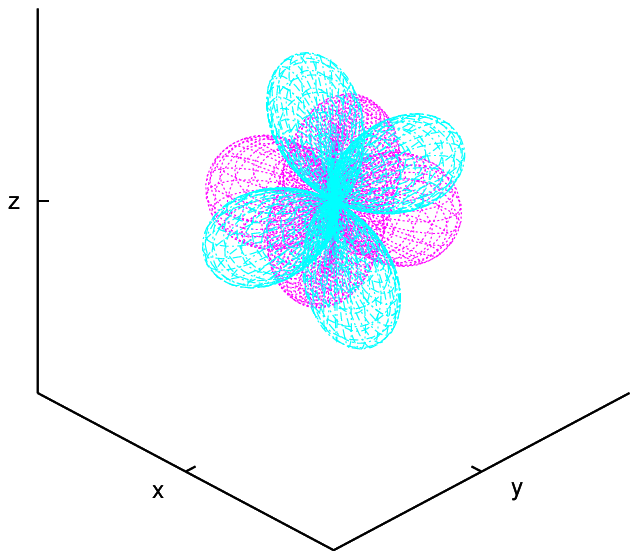}}
\caption{Symmetry adapted V($3d$) orbitals for BaVS$_3$ in the $Cmc2_1$
         structure. (a) $A_{1g}$, (b) $E_g$ as well as (c) $e_g$. In the
         orthorhombic $Cmc2_1$ structure the following correspondance for
         the axes is realized:
         $a\leftrightarrow x$, $b\leftrightarrow y$, $c\leftrightarrow z$.}
\label{dorbfig}
\end{figure}

We performed a realistic many-body investigation of the electronic
structure of BaVS$_3$ by means of the combination of the LDA with the 
DMFT. The ab-initio LDA calculations were performed with a
pseudopotential code within a mixed-basis consisting of plane waves and
localized functions~\cite{MBPP}. Norm-conserving pseudopotentials were
used, and for the exchange-correlation functional the parametrization of
Perdew and Wang \cite{Per92} was utilized. In order to account for the
given crystal symmetry of BaVS$_3$, a symmetry-adapted V(3$d$)-basis
$\{\phi_m\}$ (see Fig.\ref{dorbfig}) was obtained by diagonalizing the 
orbital density matrix $n_{MM'}$$\sim$
$\sum_{{\bf k}b}f_{{\bf k}b}\langle\psi_{{\bf k}b}|M\rangle\langle M'|
\psi_{{\bf k}b}\rangle$, where  $\psi_{{\bf k}b}$ stands for the pseudo
crystal wave function for wave vector ${\bf k}$ and band $b$, and
$M$,$M'$ denote the cubic harmonics for $\ell$=2.\\
In the LDA$+$DMFT approach the on-site Green's function of a multi-orbital
problem reads as
\begin{equation}
{\bf G}_{\sigma}(i\omega_n)=\sum_{\bf k}\left[(i\omega_n+\mu){\bf 1}-
{\bf H}({\bf k})-{\bf\Sigma}_{\sigma}(i\omega_n)\right]^{-1}
\quad,\label{green}
\end{equation}
with {\bf H}({\bf k}) as the self-consistent determined LDA hamiltonian,
provided in a localized basis, as well as the $k$-independent self-energy
${\bf \Sigma}(i\omega_n)$. In Eq. (\ref{green}), $\sigma$ is the spin
index, $\omega_n$$=$$(2n+1)\pi/\beta$ ($n=0,\pm 1,\pm 2,\ldots$) are the
fermionic Matsubara frequencies for the inverse temperature
$\beta$$\equiv$$T^{-1}$, and $\mu$ is the
chemical potential. For the special matter in hand, we restricted the
calculations to the paramagnetic case, because no magnetic order is
expected in the MIT regime. Furthermore, we approximated the full
LDA hamiltonian by its downfolded version onto the $\{A_{1g},E_g\}$ basis,
which is sufficient to capture the essential physics close to the
MIT. This downfolding to an effective 3-band model was performed
'empirically', lead by straightforward physical arguments, on the
LDA DOS~\cite{Lec05}. The V($3d$)-Green's function matrix for the 
effective 3-band model takes diagonal shape in the LDA limit. We kept 
this shape also within the DMFT framework, since off-diagonal self-energy 
terms should be small due to symmetry. Thus one may replace the $k$-sum in 
eq. (\ref{green}) by the integral over the partial DOS $D_m(\varepsilon)$, so 
that the orbital-resolved Green's functions are written as
\begin{equation}
G_m(i\omega_n)=\int\frac{d\varepsilon D^{\mbox{\tiny{(LDA)}}}_m
(\varepsilon)}{i\omega_n+\mu-\varepsilon-\Sigma_m(i\omega_n)}\quad,\quad
m=A_{1g}, E_{g1}, E_{g2}\quad.
\label{greenorb}
\end{equation}
Note that by treating only correlated states in this realistic DMFT
formalism, double-counting terms originating from correlations already
included in the LDA are absorbed by the chemical potential $\mu$. The
on-site vertex was again parametrized as
$U_{mm}^{\uparrow\downarrow}$=$U$,
$U_{m\neq m'}^{\uparrow\downarrow}$=$U$$-$$2J$ and
$U_{m\neq m'}^{\uparrow\uparrow(\downarrow\downarrow)}$=$U$$-$$3J$,
with $U$ the on-site Coulomb repulsion and $J$ the local Hund's rule
coupling. Recall that $J$ does not only
describe the spin exchange energy, but also the reduction of $U$ for
electrons in different orbitals. The local impurity problem within the
DMFT formalism was again solved using QMC with similar convergence
parameters as in the model investigations.
\subsection{The electronic structure within LDA}
\begin{figure}[b]
\parbox{\halftext}{% Fig 4: LDA band structure with fatbands
\epsfclipon
%%%%%%
\epsfig{file=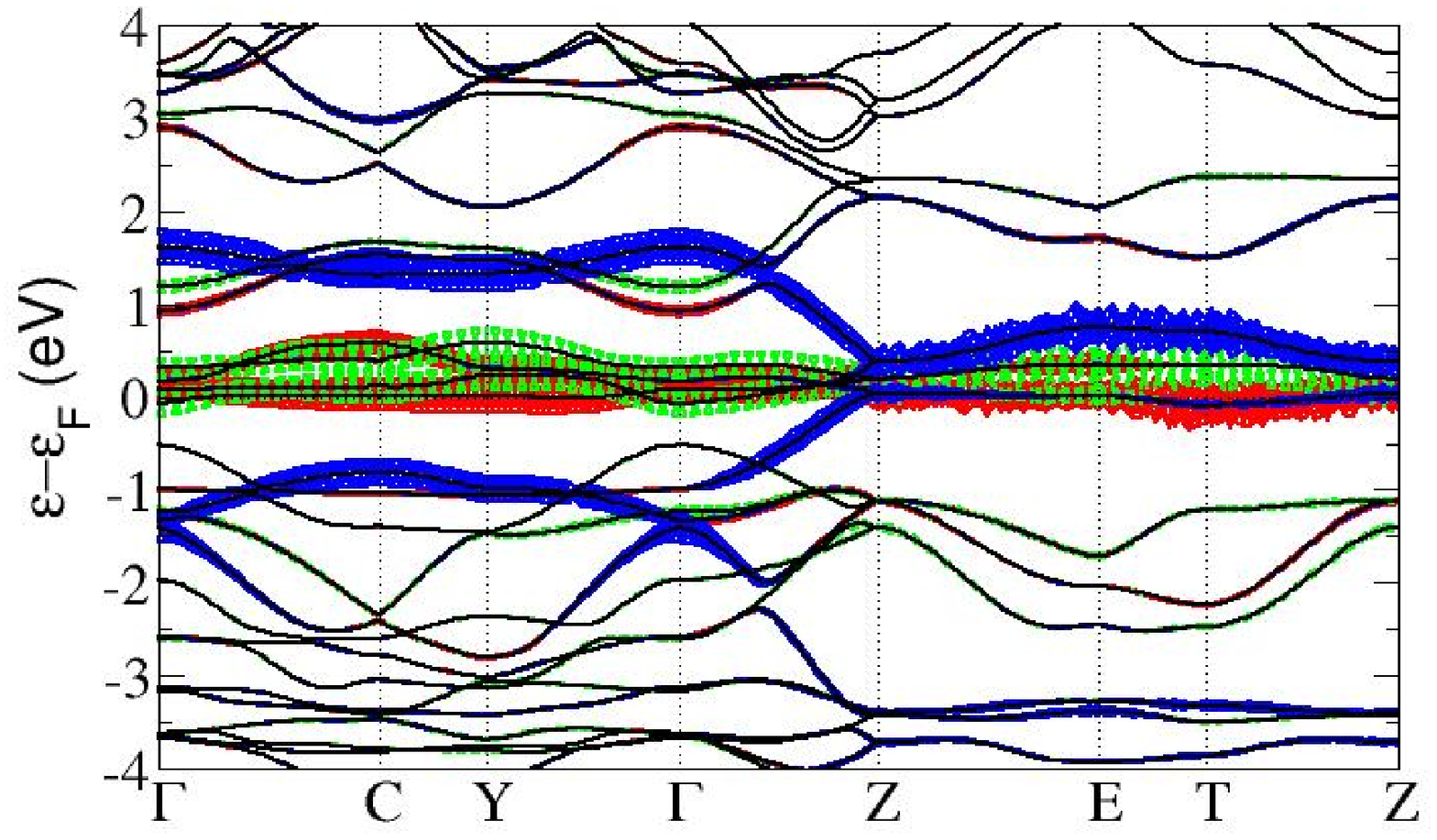,width=6.5cm}
\caption{LDA band structure for $Cmc2_1$-BaVS$_3$ 
         along high-symmetry lines in the 1.
         Brillouin Zone. Also shown are the fatbands (see text) for the
         $A_{1g}$ (blue), $E_{g1}$ (red) and $E_{g2}$ (dashed-green)
         orbital of the V atoms. The wiggling of the fatbands in the
         Z-E-T plane is due to the high degeneracy of the bands.
%{\it comment in back figure! just for quicker compilation ...}
}
\label{Fig4}}
\hfill
\parbox{\halftext}{% Fig 5: LDA partial Co-DOS
\vspace*{-1cm}
\epsfclipon
\epsfig{file=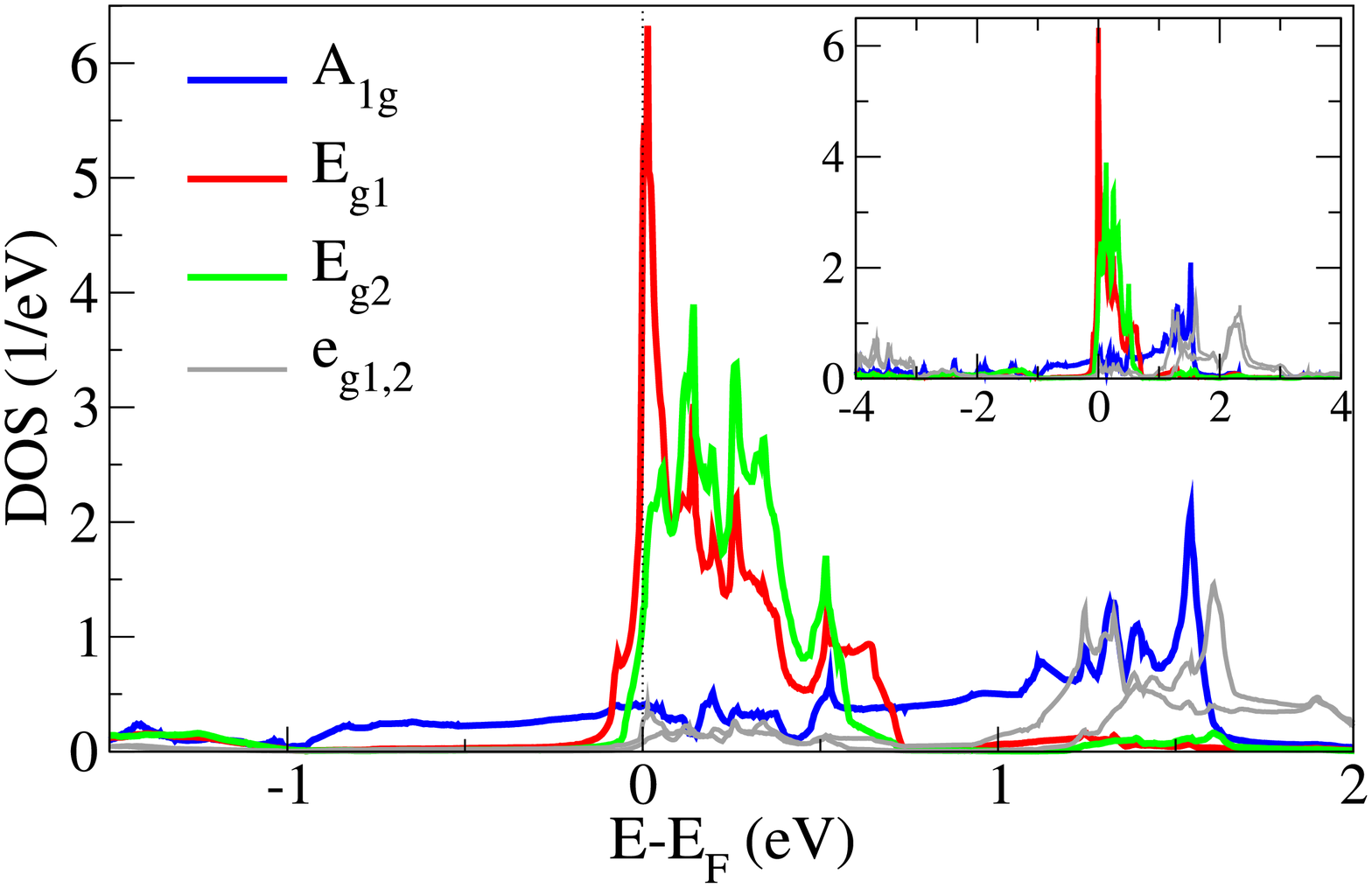,width=6.5cm}
\caption{Partial LDA-DOS of the V($3d$) states in the
         symmetry-adapted basis. For the computation of this partial DOS,
         a sphere of radius 2.18 a.u., i.e., half of the nearest-neighbor
         V-S distance, was introduced.}
\label{Fig5}}
\end{figure}

Figure \ref{Fig4} shows the computed band structure for the $Cmc2_1$
crystal data at $T$=100 K~\cite{Ghe86}. The high-symmetry points
$\Gamma$-$C$-$Y$ define a triangle in the $k_z$=0 plane, whereas 
$Z$-$E$-$T$ is the
analog shifted triangle in the $k_z$=$0.5c^*$ plane. The $\Gamma$-$Z$ line
corresponds to the propagation along the chain direction in BaVS$_3$. Also
drawn are the so-called fatbands for the $\{A_{1g},E_g\}$ orbitals. The
width of these fatbands is proportional to the amount of the orbital
character of a given band in the symmetry-adapted $\{\phi_m\}$ basis.
Being directed along the chain
direction between pairs of V atoms, the $A_{1g}$ orbital has mainly
$d_{z^2}$ character. In contrast, the $E_g$ states, linear combinations
of $d_{yz},\,d_{x^2-y^2}$ and $d_{z^2}$ ($E_{g1}$) as well as $d_{xy}$
and $d_{xz}$ ($E_{g2}$), only weakly hybridize with their surrounding.
The orbitals of the remaining $e_g$ manifold point mainly towards the
sulfur atoms, which results in a large energy splitting, leading to a
smaller(larger) contribution to the occupied(unoccupied) states well
below(above) $\varepsilon_F$. Therefore, the $e_g$ states do not have
a major influence on the essential physics around the MIT.
In Fig.~\ref{Fig4} one can clearly identify that the very narrow bands at
the Fermi level are associated with the two $E_g$ orbitals. Along
$\Gamma$-$Z$ starting at around -1 eV, one observes a nearly filled
dispersive band with mainly $A_{1g}$ character that crosses the Fermi
level close to the edge of the Brillouin zone (BZ). The $2k_F$ value for
this band amounts to 0.94$c^*$, hence is incompatible with the
experimentally observed $2k_F$=0.5$c^*$ instability. In addition, the
filling of the $E_g$ bands is rather low and cannot account for the
observed magnitude of the local moment. Note however the small $E_{g2}$
electron pocket at the $\Gamma$ point. It is absent in the band structure
of the hexagonal phase and may be related to the hole-like conduction
below $T_S$ as seen in Hall measurements~\cite{Boo99}. The relevant
$k$-summed electronic structure can be examined in Fig. \ref{Fig5} via the
$\phi_m$-resolved LDA-DOS per atom of the V($3d$) states. This plot of the
partial DOS is characterized around the energy intervall [-1,2] eV by a
rather broad $A_{1g}$ band and two strikingly narrow $E_g$ bands right at
$E_F$, in accordance with the shown band structure. The small 
crystal-field splittings read: 
$\Delta(A_{1g},E_g)$=$\varepsilon_{A_{1g}}-\varepsilon_{E_g}$$\sim$0.1 eV 
and 
$\Delta(E_{g1},E_{g2})=\varepsilon_{E_{g1}}-
\varepsilon_{E_{g2}}$$\sim$-0.05 eV.
\subsection{Orbital repopulations and Fermi surface deformations}
\begin{figure}[t]
\parbox{\halftext}{% Fig 6: Occupations and k-shift in DMFT
\epsfclipon
\epsfig{file=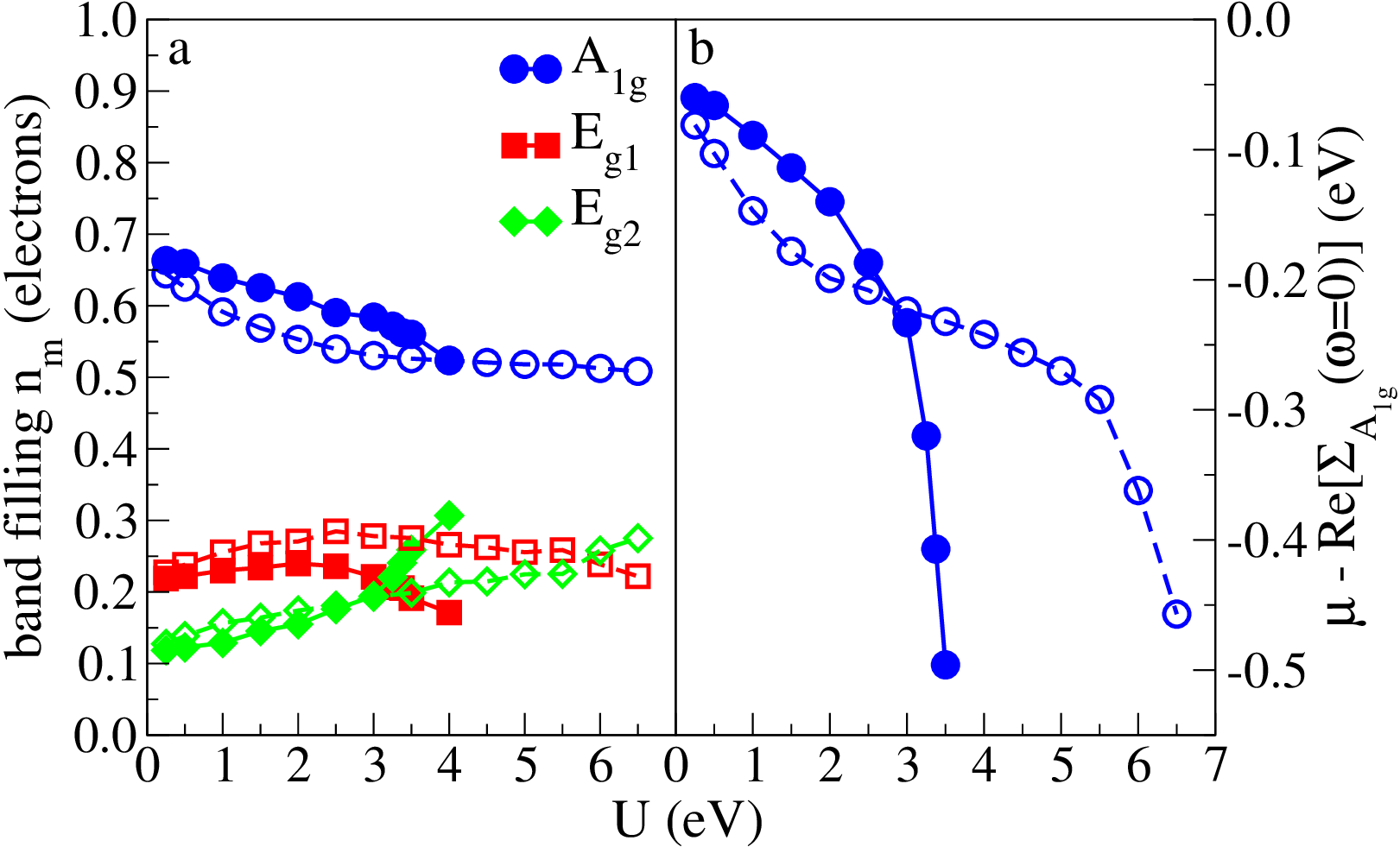,width=6.5cm}
\caption{(a) Band fillings at $\beta$=$(k_{\sub{B}}T)^{-1}$=15 eV$^{-1}$
         ($T$=$774$~K) for the effective bands within LDA+DMFT. (b)
         Corresponding shift of the Fermi level for the $A_{1g}$ band
         (note that $\varepsilon_{{\bf k}_F}^{\sub{(LDA)}}$=0). Filled
          symbols: $U/J$=7, open symbols: $U/J$=4.}
\label{Fig6}}
\hfill
\parbox{\halftext}{% Fig 7: Quasiparticles in DMFT
\epsfclipon
\epsfig{file=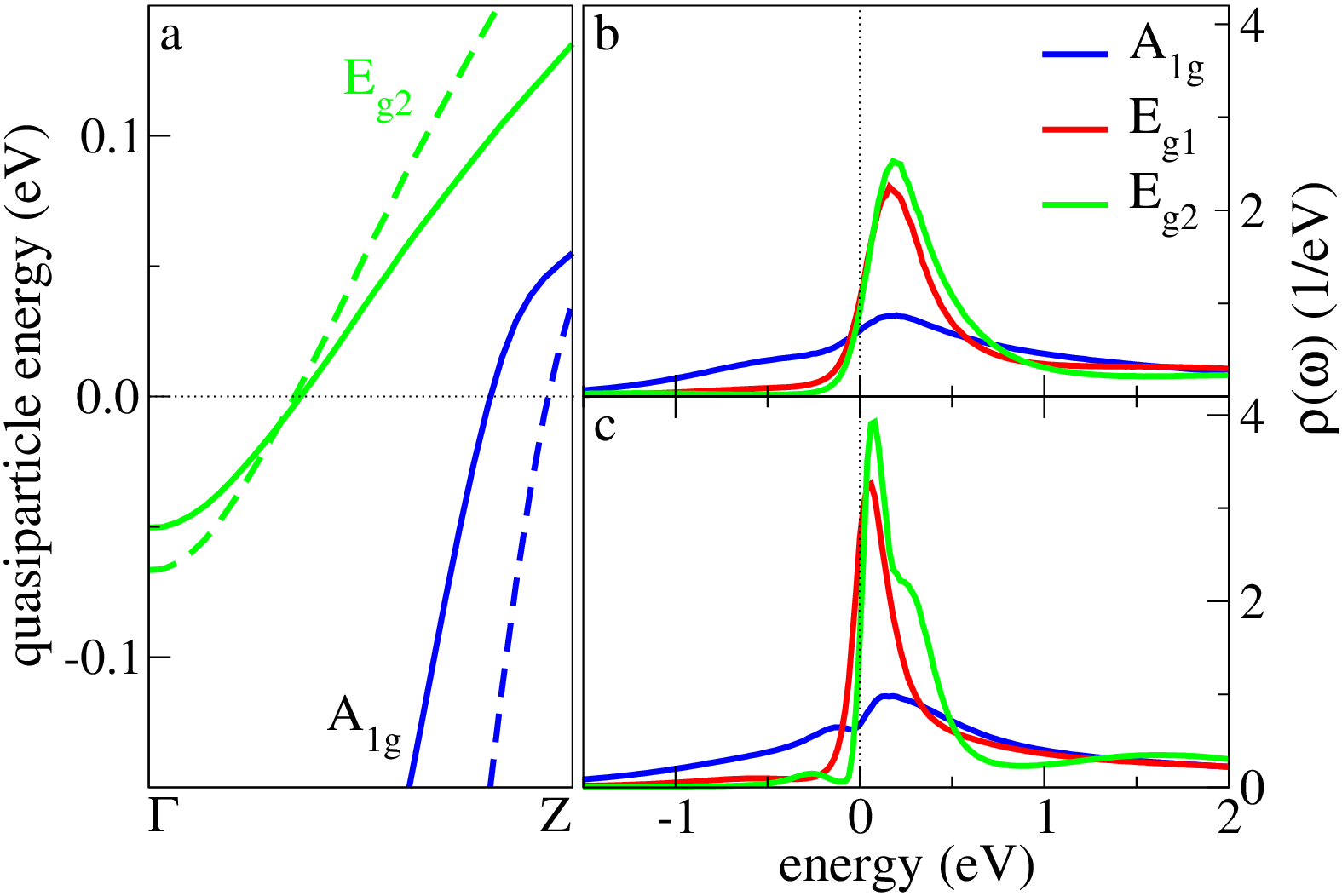,width=6.5cm}
\caption{LDA+DMFT spectral data for $U$$=$3.5 eV, $U$$/$$J$=4. (a) V($3d$)
         low-energy quasiparticle bands along $\Gamma$-$Z$ in LDA (dashed
         lines) and LDA+DMFT (solid lines) for $T$=332~K. (b,c)
         integrated spectral function $\rho(\omega)$ for a single formula
         unit of BaVS$_3$ at $T$=1160~K (b) and $T$=332~K (c).}
\label{Fig7}}
\end{figure}

We will show that the intimate interplay between the
Hubbard parameters $U$ and $J$ as well as the given LDA band
dispersions can eventually result in a substantial charge transfer from
the broader $A_{1g}$ band to the narrower $E_{g}$ bands. Additionally,
accompanied by this charge transfer is a shift of the $A_{1g}$ band in
k-space, providing the possibility for the onset of the experimentally
detected CDW instability.\\
Due to the lack of information about the value of $U$ in BaVS$_3$ from
experiment (e.g., photoemission) or theory (constrained LDA methods tend
to underestimate the screening for metals), we undertook our
investigation by carefully scanning through the physically
meaningful $U$-$J$ parameter space for this compound. Since the interplay
of $U$ and $J$ appears to be a central issue, we chose to fix the ratio
$U/J$. By varying $U$ we studied two different series: $U/J$$=$7 and
$U/J$$=$4. Recall that in the 2-band model we encountered in
\S\ref{modcalc}, the ratio $U/J$$=$5 separated the
polarization/compensation regimes in the static limit. One would expect 
to find similar regimes also in the present 3-band model for BaVS$_3$.\\
The orbital occupancies in our effective 3-band model, at the LDA level
(i.e., for $U$=$0$) read: $n(A_{1g})$=0.712, $n(E_{g1})$=0.207 and
$n(E_{g2})$=0.081. In Fig.~\ref{Fig6}a we plotted the respective
band fillings in the metallic regime for increasing $U$ within the two
$U/J$ series. The main effect apparent on Fig.~\ref{Fig6}a is that
moderate correlations indeed tend to bring the occupancies of each
orbital closer to one another, i.e., to decrease the population of the
``extended'' $A_{1g}$ orbital and to increase the occupancy of the
$E_g$ orbitals.
For strong correlations, values close to
$n(A_{1g})$$\simeq$$n(E_{g1})$$+$$n(E_{g2})$$\simeq$0.5
are obtained in the DMFT calculation, corresponding to a half-filled band.
The difference in the qualitative $U$-dependence of the band fillings
between the two $U/J$ series resembles those of the model calculations.
For $U/J$=4 there is an immediate orbital charge transfer for $U$$\neq$0,
whereas for $U/J$=7 correlations first have to overcome the polarizing
tendencies for small $U$. Hence the basic mechanisms found in the model
calculations appear to be an essential point in understanding the physics
of BaVS$_3$. Note that observations on the importance of $J$ and of
band-narrowing effects for the band filling in realistic systems have
also been pointed out in the context of ruthenates \cite{Lie00,Oka04}.
For $U$$\sim$4 eV the system reaches the Mott transition within $U/J$=7,
whereas this transition is shifted to much larger $U$ values for $U/J$=4.
In the present context, we did not investigate the Mott insulating state
as this one may not be relevant for realistic BaVS$_3$. Recall that in
nature the orthorhombic ($Cmc2_1$) structure transforms at
$T_{\sub{MIT}}$ via an CDW instability into a monoclinic ($Cm$) structure
with 4 inequivalent vanadium atoms in the unit cell. Hence an
understanding of the true insulating state cannot be obtained by driving
the system in the Mott-insulating regime. For the strongly correlated
metal close to $T_{\sub{MIT}}$ we would estimate $U$$\sim$3.5 eV in order
to explain the substantial filling of the $E_g$ states revealed from the
local moment examinations.\\
As already noted, the correlation-induced orbital charge transfers not
only modify the respective band fillings, but may lead also to relevant
changes of the Fermi surface. Such an effect is anticipated in the case
of BaVS$_3$ due to the insufficiency of DFT in LDA to explain the observed
CDW instability. Indeed our calculations reveal that the depletion of the
$A_{1g}$ band is accompanied by a reduction of the corresponding Fermi
wave vector ${\bf k}_F$ along the $\Gamma$-$Z$ direction. While a full
determination of the quasiparticle (QP) band structure in the interacting
system requires a determination of the real-frequency self-energy, we can
extract the low-energy expansion of this quantity from our QMC
calculation in the form:
$\mbox{Re}\Sigma_m(\omega+i0^+)$$\simeq$$
\mbox{Re}\Sigma_m(0)+\omega(1-1/Z_m)+\cdots$, with $Z_m$ the QP
residue associated with each orbital. The poles of the
Green's function determine the QP dispersion relation:
$\mbox{det}[\omega_{\bf k}-\hat{Z}[\hat{H}_{\bf k}^{\rm{LDA}}+
\mbox{Re}\hat{\Sigma}(0)-\mu]]$=0,
with $\mu$ the chemical potential. Focusing first on the $A_{1g}$ sheet of
the Fermi surface, within our diagonal formulation the location of the
Fermi wave vector in the interacting system is determined by:
$\varepsilon^{\rm{LDA}}_{A_{1g}}({\bf k}_F)$=
$\mu-\mbox{Re}\Sigma_{A_{1g}}(0)$.
This quantity therefore yields the energy shift of the $A_{1g}$ band at
the Fermi surface crossing, as compared to LDA. It is depicted in
Fig.~\ref{Fig6}b as a function of $U$. In Fig.~\ref{Fig7}a, we display
the QP bands that cross the Fermi level along $\Gamma$-$Z$ in a narrow
energy range around $\varepsilon_F$ for $U$=3.5 eV in the $U/J$=4 series.
The QP bands are obtained by performing a perturbative expansion of the
pole equation above, which yields:
$\omega_{b{\bf k}}$=
$\sum_m C_{m{\bf k}}^{b}Z_m[\varepsilon_{b{\bf k}}^{\rm{LDA}}
+\mbox{Re}\Sigma_m(0)-\mu]$ with
$C_{m{\bf k}}^{b}$$\equiv$$|\langle\psi_{{\bf k}b}|\phi_m\rangle|^2$
the LDA orbital weight. From these two figures, it is clear that
$k_F(A_{1g})$ is reduced in comparison to the LDA value, in line with the
global charge transfer from $A_{1g}$ to $E_g$. This opens new
possibilities for the CDW instability, in particular for the nesting wave
vector.\\
\begin{figure}[t]
\begin{center}
% Fig 8: local susceptibilities
\epsfclipon
\epsfig{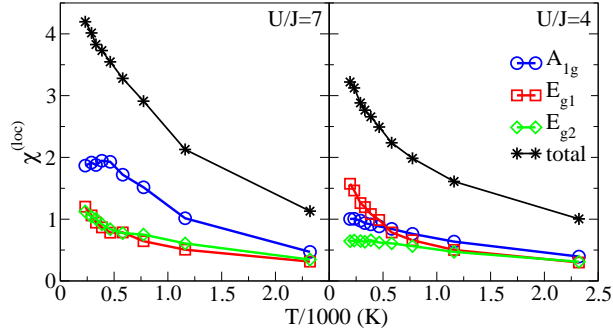}
\caption{Temperature-dependent local spin susceptibilities for
         $U$=3.5 eV, according to the normalization
         $\chi^{\sub{(loc)}}$=$\int_0^\beta d\tau\langle
         \hat{S}_z(0)\hat{S}_z(\tau)\rangle$, where $\hat{S}_z$ denotes
         the $z$-component of the spin operator.}
\label{Fig8}
\end{center}
\end{figure}
The enhanced population of the narrow $E_g$ bands, as well as the
correlation-induced reduction of its bandwidth (see Fig.~\ref{Fig7}a)
provide an explanation for the local moments observed in the metallic
phase. To support this, we have calculated (Fig.~\ref{Fig8})
the local susceptibility associated with each orbital
$\chi^{\rm{(loc)}}_m\equiv\sum_{{\bf q}}$Re$[\chi_m({\bf q},\omega$=0)].
For both values of $U/J$, the susceptibility of the $A_{1g}$ band
saturates to a Pauli-like value at low temperatures. In contrast,
$\chi^{\rm{(loc)}}$ of the $E_g$ orbitals strongly increases as the
temperature is lowered (except for the low-filled $E_{g2}$ orbital at
$U/J$=4). This is because the coherence temperature below which
quasiparticles form is much lower for the $E_g$ orbitals than for the
$A_{1g}$ orbital. Accordingly, our calculation of the integrated spectral
functions (Fig.~\ref{Fig7}b,c) reveals a strong $T$-dependence of the
$E_g$ QP peak. Some differences between the two series are clear from
Fig.~\ref{Fig8}. For $U/J$=7, the system is already very close to the
Mott transition when choosing $U$=3.5 eV. Thus the $A_{1g}$ electrons
also act as local moments over part of the temperature range, while for
$U/J$=4 and same $U$, the $T$-dependence of the total local
susceptibility is almost entirely due to the $E_{g1}$ electrons.
Which of the two situations is closest to the physics of BaVS$_3$
does require further investigations, albeit some experimental indications
point at the second possibility~\cite{Faz02}.

\section{The Fermi surface of Na$_x$CoO$_2$}
Since the surprising discovery of superconductivity in
Na$_x$CoO$_2$$\cdot y$H$_2$O~\cite{Tak03} for $x$$\sim$0.35, the sodium
cobaltate system has made its way to one of the hottest topics in
current research on strongly correlated materials. The layered
CoO$_2$ oxides, held together by Na layers in between, resemble the
famous cuprates, although the Co atoms form a triangular lattice
with the oxygen atoms in a pyramidal configuration above and below the
Co plane. Besides the superconducting phase the unhydrated sodium
cobaltates exhibit additional rich physics, such as strong electric
thermopower~\cite{Ter97}, charge order~\cite{Muk04}, different
magnetic order~\cite{Luo04}, as well as complex ordering of the Na
atoms~\cite{Zan04}. Bounded by a Mott-Hubbard insulator for $x$=0, i.e.,
CoO$_2$, and by the band insulator NaCoO$_2$ for $x$=1, the sodium
cobaltate phase diagram~\cite{Foo04} is roughly separated in a Pauli-like
metallic region for $x$$<$0.5 and a Curie-Weiss-like metallic region for
$x$$>$0.5. In another difference to the cuprates, cobaltate is a
multi-orbital $3d^{5+x}$ system. Due to the crystal-field splitting the Co 
atoms are in a low-spin state, hence the $d$ electrons occupy the lower 
$t_{2g}$ band complex (i.e., there are $1-x$ holes in the t$_{2g}$ shell). 
It follows that the Curie-Weiss behavior stems from Co$^{4+}$
local moments, whereas these moments seem to disappear below $x$=0.5.
Interestingly, magnetic phases are experimentally verified for
$x$$\ge$0.75~\cite{Luo04}.

From this short introduction of Na$_x$CoO$_2$ it is 
already obvious that the $x$=0.5 composition plays a subtle role
in this system. Although the characteristics of the phase at $x=$0.5 are
still not completely resolved, certain properties seem to be established.
In contrast to most other doping levels $x$, Na$_{0.5}$CoO$_2$ exhibits
an ordered superstructure for the combined Na/CoO$_2$ system~\cite{Hua04}
over a wide temperature range. Most importantly, Na$_{0.5}$CoO$_2$ shows
a MIT at $\sim$53 K. If this transition is truly associated with the
onset of charge ordering is still a matter of debate, although many
measurements point in this direction~\cite{Foo04,Wan04}. There appear to
be~\cite{Hua04} additional magnetic and/or structural transitions at 20 K
and 87 K.

Early electronic structure
calculations in the L(S)DA framework for Na$_{0.5}$CoO$_2$ have been
performed by Singh~\cite{Sin00}. Since the the ordered superstructure was
identified only recently, in these DFT calculations a disordered
arrangement of the sodium atoms was assumed (treated within the virtual
crystal approximation). The Fermi surface revealed in this investigation
exhibits a large cylindrical hole sheet with dominant $A_{1g}$ character
around the $\Gamma$-$A$ line in the Brillouin zone of the hexagonal unit
cell, as well as small hole pockets in the $\Gamma$-$K$ and $A$-$H$ 
directions. This
topology of the Fermi surface appears to be rather generic for a wide
range of $x$ within the LDA framework. It has been
suggested~\cite{Joh04} that strong nesting between the hole pockets leads
to large fluctuations in the spin channel, having important
influence on the character of the superconducting state at $x$$\sim$0.35.
However the existence of these hole pockets has not been confirmed in
angle-resolved photoemission (ARPES) experiments~\cite{Has04,Yan05,Gec05}. On
the contrary, in one of these ARPES measurements~\cite{Yan05} the
``pocket'' bands were found well below the Fermi level for large doping
range $x$.

It appears obvious from the latter discussion that Fermi-surface
modifications due to strong correlation effects may play also an
important role for sodium cobaltates. Hence the last
sections are devoted to a brief analysis of the problem of finding the
correct generic Fermi surface for Na$_x$CoO$_2$.
\subsection{Electronic structure within LDA}
\begin{figure}[b] % Fig9 : Na0.5CoO2 crystal strcuture
\parbox[t]{4cm}{(b)\\[0.2cm]
\epsfxsize=0.4\textwidth
\epsfbox{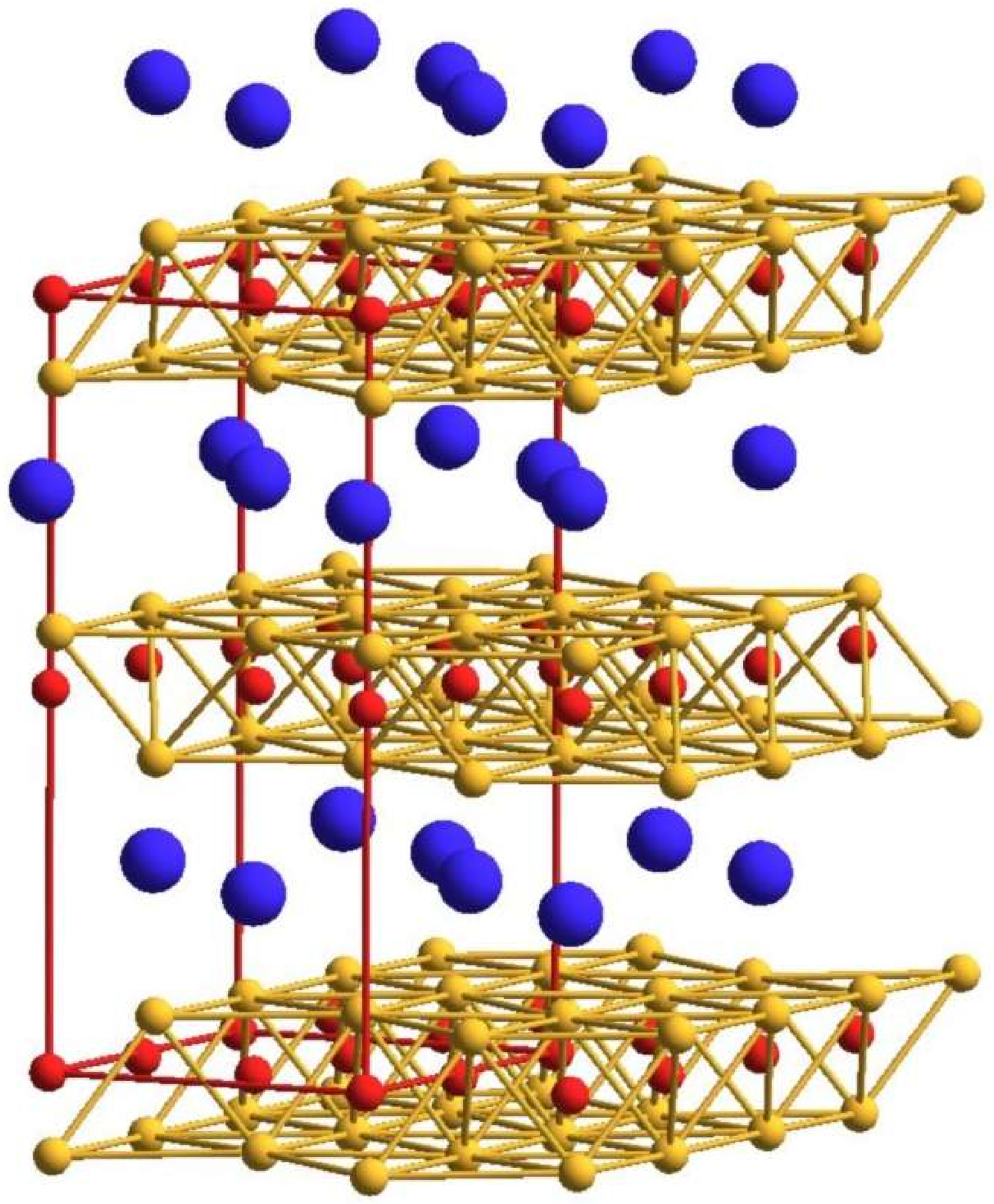}}
\parbox{1.75cm}{\hfill}
\parbox[t]{7cm}{(b)\\[0.2cm]
\epsfxsize=0.4\textwidth
\epsfbox{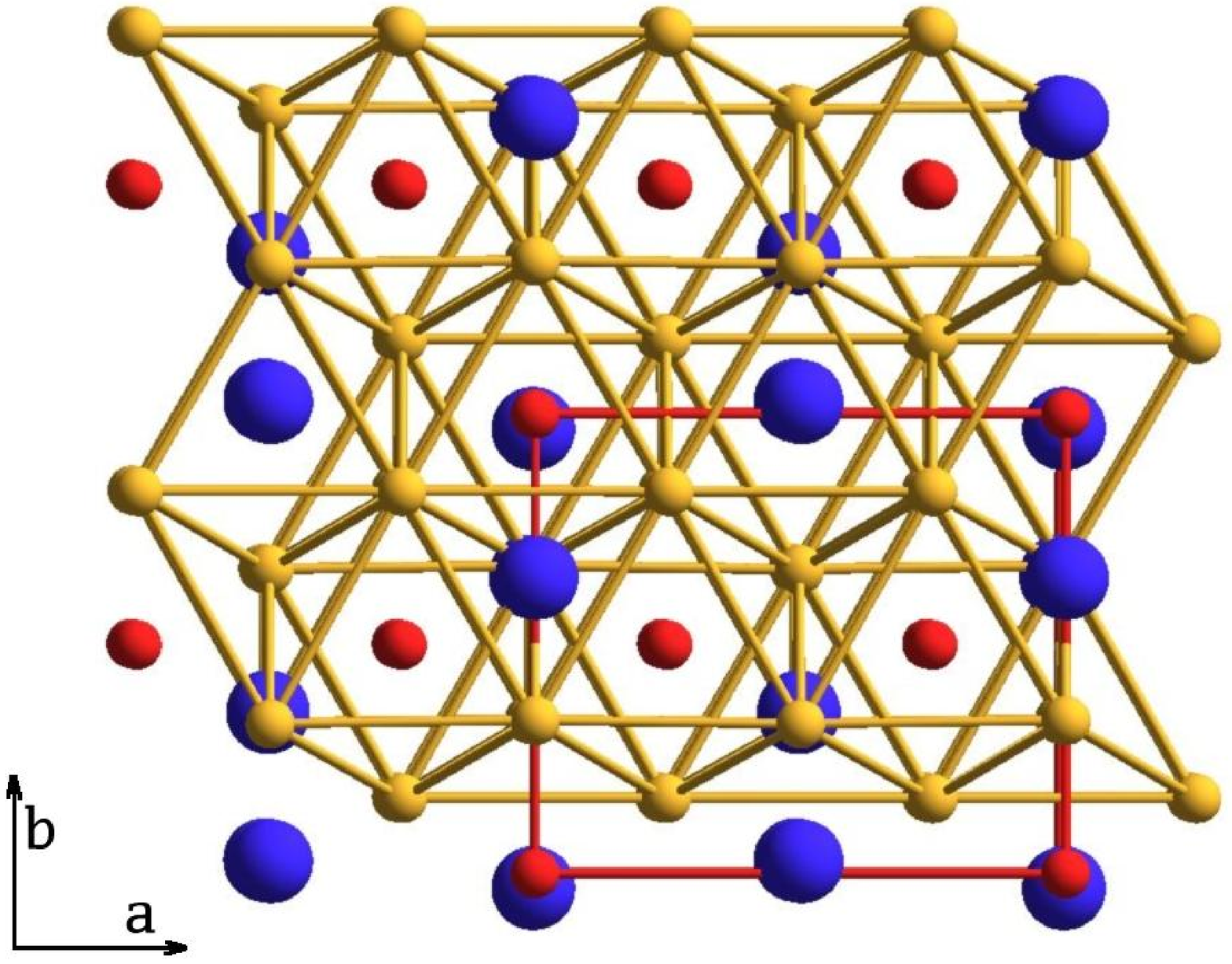}\\
\epsfxsize=0.1\textwidth
\hspace{0.25cm}\epsfbox{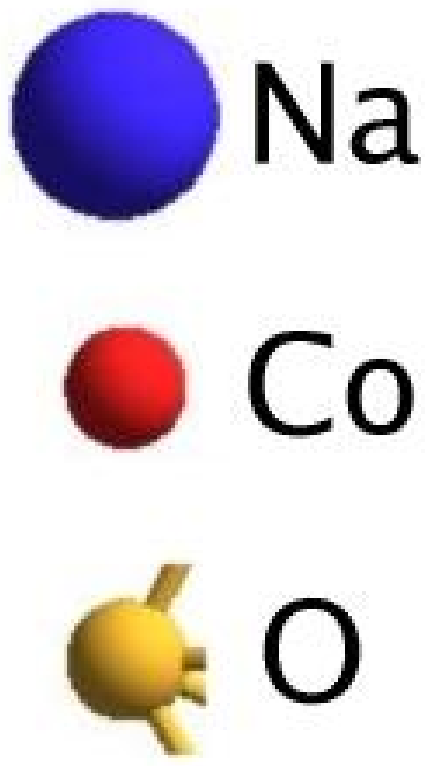}}
\caption{Na$_{0.5}$CoO$_2$ in the orthorhombic ($Pmmn$) structure.
(a) 3D view, (b) view along the $c$ axis.}
\label{Fig9}
\end{figure}
In order to have a well-defined starting point, we tried to find
theoretical agreement with the experimentally detected ordered
superstructure for Na$_{0.5}$CoO$_2$. Thus we performed total energy
calculations with the mixed-basis
pseudopotential (MBPP) code for different arrangements of sodium atoms in
the interlayers in between the CoO$_2$ planes. As a starting point for the
Na$_x$CoO$_2$ structural model one usually utilizes the hexagonal
($P6_322$ or $P6_3/mmc$) symmetry \cite{Jan74,Sin00,Hua04}. In all our
studies the lattice parameters were confined to the values obtained by
Jansen and Hoppe~\cite{Jan74}, i.e., $a_H$=2.84\AA and $c_H$=10.81\AA.
Indeed we obtained the lowest LDA total energy for the experimentally
suggested superstructure with global orthorhombic ($Pmmn$)
symmetry~\cite{Hua04}. We relaxed the atomic positions within the MBPP and
the corresponding superlattice is depicted in
Fig.~\ref{Fig9}. The unit cell of this superstructure with dimension
$a_O$=$2a_H$,$b_O$=$a_H\sqrt{3}$ and $c_O$=$c_H$ contains 28 atoms, 
whereby there are 2
inequivalent Na and Co sites, respectively, as well as 3 inequivalent O
sites. The two classes of inequivalent Na and Co sites are related as the
Na ordering is such that in each sodium interlayer there are atoms
sitting on top of an Co site (Na1/Co1) and such atoms that are placed on
top of an oxygen site (Na2) (see. Fig.~\ref{Fig9}b). The Co atoms without
an Na atom on top form the Co2 class. As reported by Huang {\sl et al.}
\cite{Hua04}, this ordered superstructure should be stable over a wide
temperature range, with possible structural changes below 100 K.

\begin{figure}[t]
\parbox{\halftext}{% Fig 10: LDA band structure with fatbands
\epsfclipon
\epsfig{file=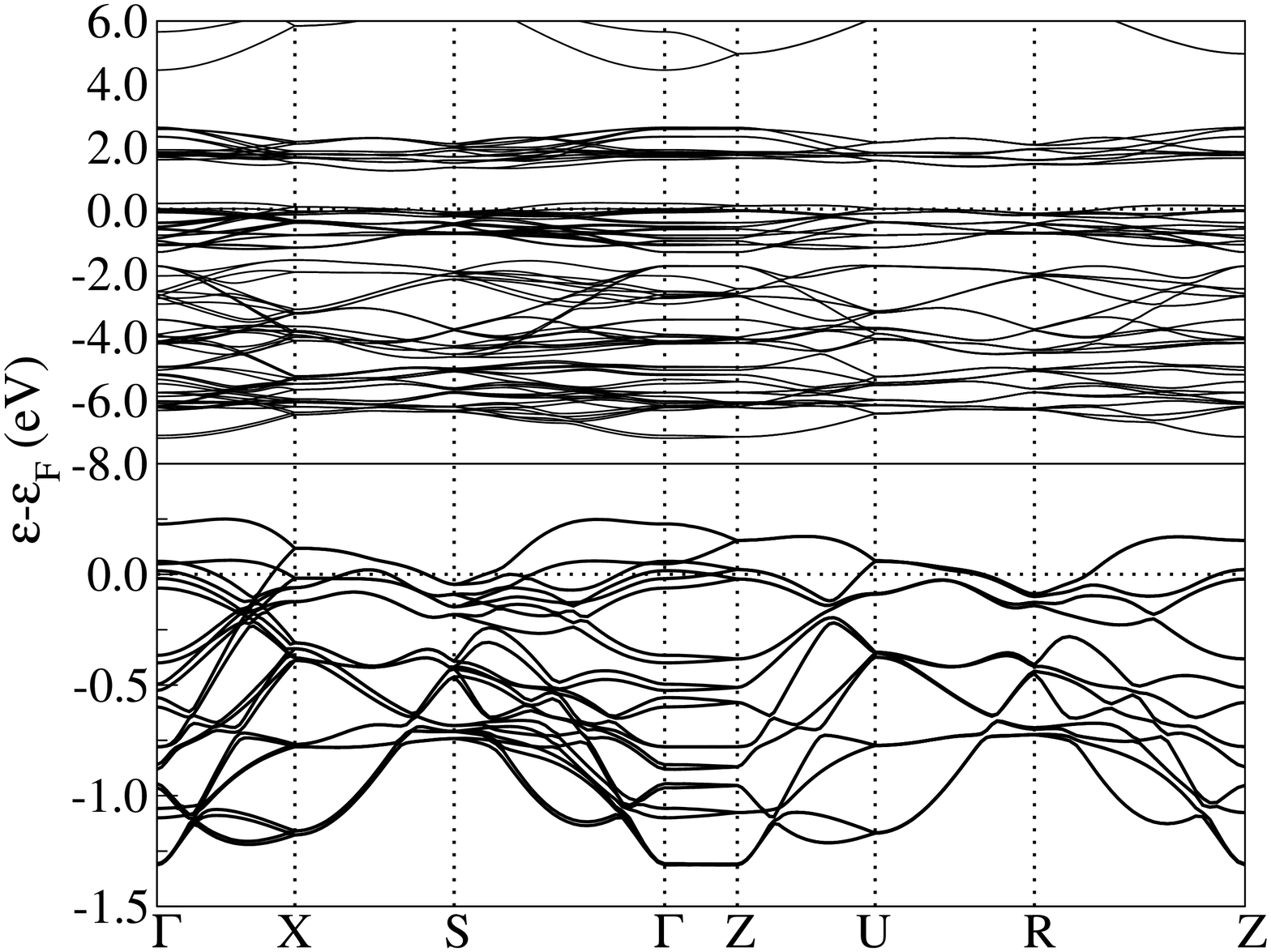,width=6.5cm}
\caption{LDA band structure for Na$_{0.5}$CoO$_2$ in the $Pmmn$ crystal
         structure along high-symmetry lines in the 1. BZ of the
         orthorhombic lattice.}
\label{Fig10}}
\hfill
\parbox{\halftext}{% Fig 11: LDA partial V-DOS
\epsfclipon
\epsfig{file=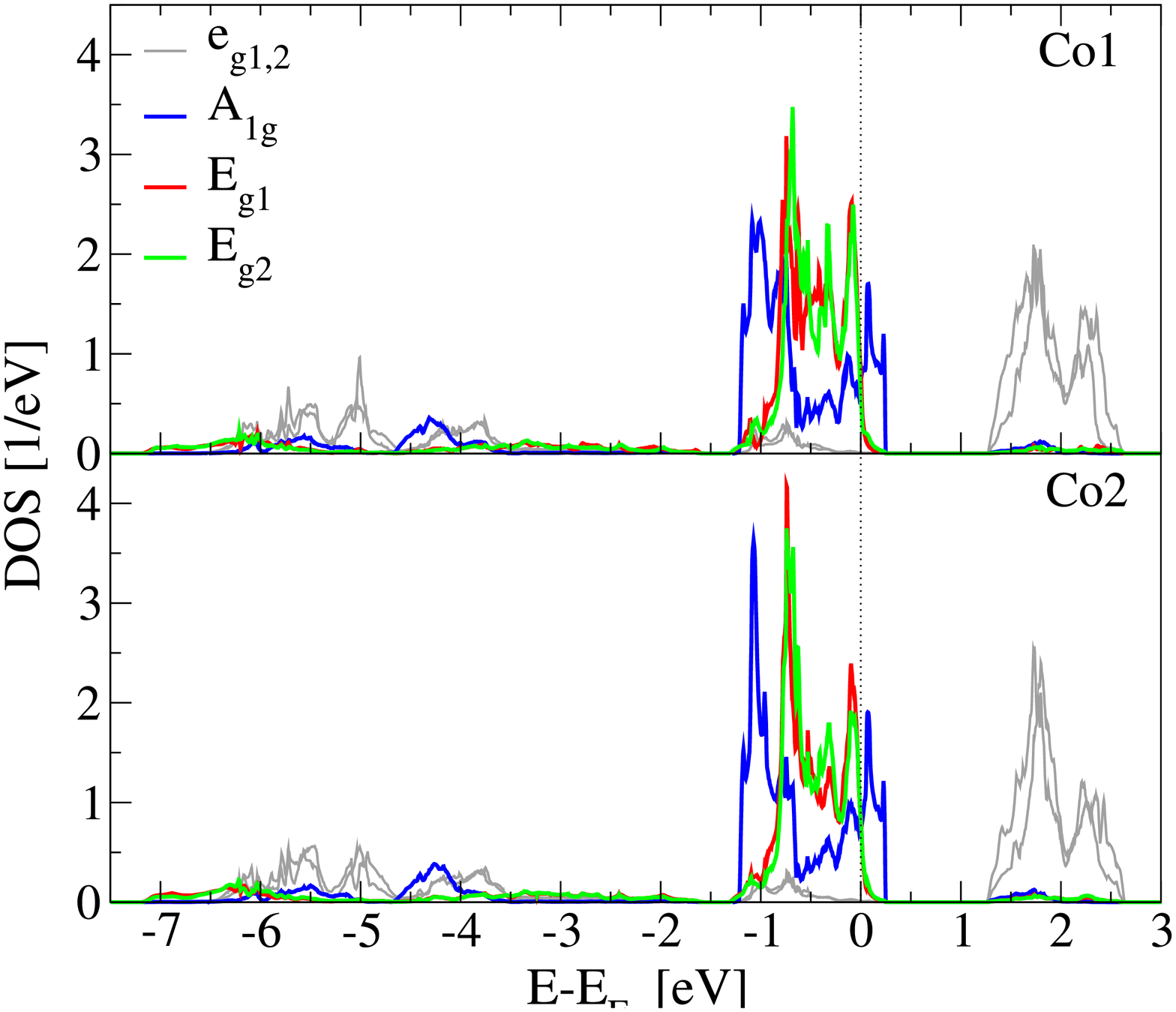,width=6.5cm}
\caption{Partial LDA-DOS of the Co$(3d)$ states in the symmetry-adapted
         basis for Co1 and Co2 atoms.}
\label{Fig11}}
\end{figure}
\begin{figure}[b]
\parbox{\halftext}{% Fig 12: bands in hexgonal Brillouin Zone
\epsfclipon
\epsfig{file=na05coo2.hexbands.eps,width=7cm}
\caption{LDA band structure for Na$_{0.5}$CoO$_2$ in the $Pmmn$ crystal
         structure along high-symmetry lines in the 1. BZ of the
         hexagonal lattice with $a$=$a_H$.}
\label{Fig12}}
\hfill
\parbox{\halftext}{% Fig 13: fatbands in hexagonal BZ
\epsfclipon
\epsfig{file=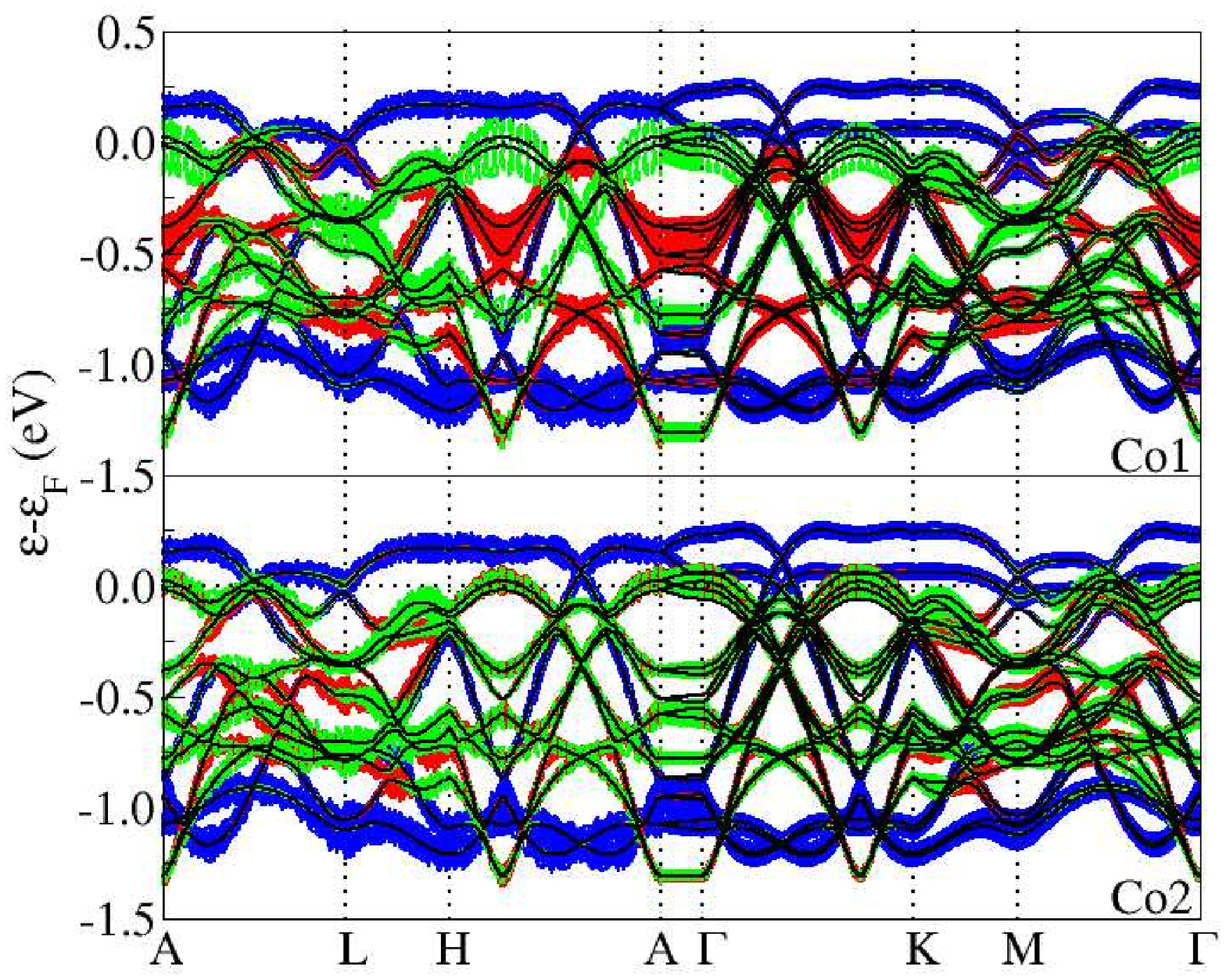,width=6cm}
\caption{$t_{2g}$ fatbands for the Co1 and Co2 class in the hexagonal BZ
         ($A_{1g}$:blue, $E_{g1}$:red and $E_{g2}$:green).}
\label{Fig13}}
\end{figure}
We determined the LDA electronic structure for the $Pmmn$ symmetry with
the MBPP code (a similar study was published recently~\cite{Li05}). The 
resulting band structure
and the summetry-adapted $3d$-DOS for the two Co classes are shown in
Fig.~\ref{Fig10} and Fig.~\ref{Fig11}. The band structure around the Fermi
level consists of 3 isolated blocks.
Roughly speaking, the block deep in energy is formed by the O$(2p)$ states
and the one high in energy stems from the Co($e_g$) states. The most
interesting block of bands close to the Fermi level is mainly dominated by
the Co($t_{2g}$) multiplet and has only minor O($2p$) weight due to the
remaining Co($t_{2g}$)-O($2p$) hybridization. Thereby, the $A_{1g}$
subband complex
is characterized by an substantial bonding-antibonding splitting, in
between which the $E_g$ subband complex is mainly located. A crystal-field
splitting of $\Delta$=$\varepsilon_{A_{1g}}-\varepsilon_{E_g}\sim$-0.1 eV
between these two band complexes can be derived. The crystal-field
splitting between the different $E_g$ states appears to be negligible.
From Fig.~\ref{Fig11} no striking difference between the electronic states
within the named $t_{2g}$ multiplet for the two Co classes, i.e., Co1 and
Co2, can be extracted. For the respective LDA orbital occupations
the following inequalities hold: $n_{t_{2g}}^{(Co1)}>$$n_{t_{2g}}^{(Co2)}$
and $n_{E_{g1},E_{g2}}$$>$$n_{A_{1g}}$. Hence the two Co classes are
discriminated in LDA by a minor charge disproportionation
($\sim$0.06 electrons), and the results support the suggestion~\cite{Hua04}
that the Co1 atoms should tend to the lower oxidation
state, i.e., Co$^{3+}$ rather than Co$^{4+}$. However in this work we do
not address the question of charge ordering.
Finally, the hole occupancy is enlarged in the $A_{1g}$ orbital for both 
Co classes.\\
For the discussion of the LDA Fermi surface we plotted
in Fig.~\ref{Fig12} also the band structure close to the Fermi level
for the larger ($a$=$a_H$) hexagonal BZ (with symmetry lines in direct
comparison to Singh's work~\cite{Sin00}), since the generic
Na$_x$CoO$_2$ system is most often associated with the hexagonal symmetry
stemming from the CoO$_2$ sublattice. In this representation
higher BZs of the orthorhombic structure are touched, thus the band
stucture exhibits some folding. We also provide the fatbands for
the respective $t_{2g}$ orbitals in Fig.~\ref{Fig13}. By comparing the
band structure of the ordered superstructure in Fig.~\ref{Fig12} with the
one assuming disordered sodium interlayers in a smaller unit cell from
Ref.~\citen{Sin00}, some differences may be identified. There are
bands in the $A$-$\Gamma$ direction right at the Fermi level which result 
in an additional Fermi surface sheet for the orthorhombic structure. ARPES
intensity
close to the $\Gamma$ point has been detected recently but below the Fermi
level~\cite{Yan05}. In the same ARPES study the famous pocket bands were
found {\it below} the Fermi level, whereas in our LDA calculation they 
form, of course, still hole Fermi sheets. It is instructive to
study the main orbital character of the relevant bands within the
$t_{2g}$ block. From Fig.~\ref{Fig13}, the hole pockets
are nearly exclusively stemming from the $E_g$ orbitals of Co. The same
holds for the bands along $A$-$\Gamma$ at the Fermi level. As
a general subtle difference between Co1 and Co2, one observes that in the
case of Co1 the $E_g$ bands have either $E_{g1}$ or $E_{g2}$ character,
whereas for Co2 these bands have mixed $E_{g1,2}$ character. Since the
large hole Fermi sheet has dominant $A_{1g}$ character ARPES measurements
suggest a strong orbital polarization for Na$_{0.5}$CoO$_2$, namely with 
nearly entirely filled $E_g$ bands and $A_{1g}$ hole bands with a filling 
close to 0.5 holes per Co atom.

\subsection{Pockets or no pockets?}
Since the $t_{2g}$-manifold bandwidth $W$ is very narrow, Na$_x$CoO$_2$ is
expected to be in the regime of strong correlations $W$$\ll$$U$. This is
qualitatively consistent with the substantial mass renormalization and
high thermopower. A value of $U$$\approx$4 eV has been 
estimated~\cite{Joh204}. Hence, considerations from DFT-LDA may not 
provide the correct picture of the electronic structure, even for the 
Fermi surface itself.
In a recent LDA+U calculation~\cite{Zha04} the pocket bands were shifted
below the Fermi energy for various dopings $x$ (corresponding to increased
orbital polarization), in agreement with the ARPES measurements.
We have also performed (unpublished) LDA+U calculations of orthorhombic
Na$_{0.5}$CoO$_2$ and confirm this result. However, the static LDA+U 
approach is certainly questionable for the metallic regime of 
Na$_{0.5}$CoO$_2$. Moreover, the precise shape of the Fermi surface in 
LDA+U depends on the specific choice of the double-counting correction 
terms. Recently, an LDA+DMFT calculation~\cite{Ish04} obtained
the opposite effect for realistic values of $J$, namely a further 
stabilization of the hole pockets compared to LDA. In contrast, another 
study based on the Gutzwiller approximation reported that the hole pockets
were shifted below the Fermi level~\cite{Zho05}, but this is perhaps not 
surprising since $U$$\rightarrow$$\infty$ and $J$=0 were used in this 
calculation, hence favoring orbital polarization. Note that the sodium
cobaltates have non-integer total hole filling $(1-x)$$\in$[0,1] per Co
atom. As explored in \S\ref{modcalc}, the direction of the charge flow 
leading to orbital compensation or polarization is the result of a delicate
balance between different parameters, namely: the total filling, the 
Hund's rule coupling, the crystal-field splitting, and the bandwidth of 
each band (taking into account the renormalization of the latter two by 
strong correlations).
Further studies are therefore necessary to fully settle the issue for 
Na$_{x}$CoO$_2$. To conclude, let us emphasize that the following aspects,
neglected in previous studies, might be important in order to reach this 
goal:
\begin{itemize}
\item Non-diagonal terms $\Sigma_{mm'}$ in the self-energy, renormalizing 
      the interorbital hybridizations and therefore influencing charge 
      flows.
\item Spin-flip and pair-hopping (non-Ising) terms in the Coulomb vertex.
\item Importantly, an {\it orbital-independent} double-counting correction
      within the t$_{2g}$ multiplet has usually been assumed when
      including dynamical correlations. While orbital-dependence of 
      this correction is expected to be rather small and hence should not 
      affect significantly our conclusions for the larger charge-transfer 
      effects in the case of BaVS$_3$, it may play a substantial role in 
      the more delicate case of Na$_{x}$CoO$_2$.
\item Finally, non-local effects might be important, such as those induced
      by a strong interatomic Coulomb interaction. The latter have been 
      proposed~\cite{Mot04} to be important for the cobaltates, as also  
      suggested by the observed tendency to charge ordering.
\end{itemize}

\section*{Acknowledgements}
We would like to acknowledge useful discussions with J.W.~Allen, H.~Alloul,
S.~Fagot, P.~Fazekas, L.~Forr{\'o}, P.~Foury-Leylekian, J.~Geck, 
M.D.~Johannes, P.A.~Lee, A.~Liebsch, A.~Millis, J.-P.~Pouget and S.~Ravy. 
This work was supported by \'{E}cole Polytechnique and CNRS, as well as by the 
``Psi-k $f$-electron'' Network (HPRN-CT-2002-00295) and a grant of 
supercomputing time at IDRIS Orsay (project 051393).

%\clearpage

\newpage

\end{document}